\newif\ifarxiv
\arxivtrue
\ifarxiv
\documentclass[format=acmsmall, nonacm=true, timestamp=true]{acmart}
\else
\documentclass[format=acmsmall, authordraft=false]{acmart}
\fi

\ifarxiv
\pdfoutput=1 
\else
\fi

\usepackage{amsthm}
\usepackage{stmaryrd}
\AtEndPreamble{\usepackage{laws}}

\newcommand{\draftonly}[1]{}

\usepackage{imakeidx}
\makeindex

\newcommand{\fun}{\mathrel{\rightarrow}}
\newcommand{\refsto}{\mathrel{\succeq}}

\newcommand{\nondet}{\mathbin{\vee}}
\newcommand{\Nondet}{\mathbin{\textstyle\bigvee}}
\newcommand{\meet}{\mathbin{\wedge}}
\newcommand{\Meet}{\mathop{\bigwedge}}
\newcommand{\Seq}{\mathbin{;}}
\newcommand{\together}{\mathbin{\Cap}}

\newcommand{\Nil}{\boldsymbol{\tau}}
\newcommand{\cgd}[1]{\mathop{\tau}#1}
\newcommand{\pstepd}{\pi}
\newcommand{\estepd}{\epsilon}

\newcommand{\cstepd}{\boldsymbol{\alpha}}
\newcommand{\cstep}[1]{\mathop{\alpha}#1}
\newcommand{\cpstepd}{\boldsymbol{\pstepd}}
\newcommand{\cpstep}[1]{\mathop{\pstepd}#1}
\newcommand{\cestepd}{\boldsymbol{\estepd}}
\newcommand{\cestep}[1]{\mathop{\estepd}#1}
\newcommand{\Abort}{\top}
\newcommand{\Magic}{\bot}

\newcommand{\ptran}[2]{#1 \xrightarrow{\pi} #2}
\newcommand{\etran}[2]{#1 \xrightarrow{\epsilon} #2}
\newcommand{\ptranssp}{\ptran{\sigma}{\sigma'}}
\newcommand{\etranssp}{\etran{\sigma}{\sigma'}}

\newcommand{\kw}[1]{\mathbf{#1}}

\newcommand{\Do}{\mathbin{\kw{do}}}
\newcommand{\Else}{\mathbin{\kw{else}}}
\newcommand{\lblot}{\langle\mkern -3.5mu{|}}
\newcommand{\rblot}{|\mkern -3.5mu{\rangle}}
\newcommand{\Ex}[1]{{\color{blue}#1}}
\newcommand{\LEx}[1]{{\color{purple}#1}}
\newcommand{\Expr}[2]{\Ex{\lblot#1\rblot_{#2}}}
\newcommand{\LExpr}[2]{\LEx{\lblot#1\rblot_{#2}}}
\newcommand{\Eval}[2]{\Ex{\llbracket#1\rrbracket}_{#2}}
\newcommand{\LEval}[2]{\LEx{\llbracket#1\rrbracket}_{#2}}
\newcommand{\Const}[1]{\Ex{#1}}
\newcommand{\Unary}[2]{\Ex{\mathop{#1}#2}}
\newcommand{\Binary}[3]{\Ex{#1\mathbin{#2}#3}}
\newcommand{\Deref}[1]{\Ex{*\LEx{#1}}}
\newcommand{\Variable}[1]{\LEx{#1}}
\newcommand{\Base}[1]{\LEx{#1}}
\newcommand{\Array}[2]{\LEx{#1[\Ex{#2}]}}
\newcommand{\Indexed}[2]{\LEx{#1\left[\Ex{#2}\right]}}
\newcommand{\UnaryOp}{\mathop{\ominus}}
\newcommand{\BinaryOp}{\mathop{\oplus}}
\newcommand{\Fi}{\mathop{\kw{fi}}}
\newcommand{\Guarantee}{\mathop{\kw{guar}_{\boldsymbol{\pi}}}}
\newcommand{\Assign}{\mathrel{:=}}
\newcommand{\Assignment}[2]{\LEx{#1}\Assign\Ex{#2}}
\newcommand{\If}{\mathop{\kw{if}}}

\newcommand{\Od}{\mathbin{\kw{od}}{}}

\newcommand{\Rely}{\mathop{\kw{rely}}}

\newcommand{\Then}{\mathbin{\kw{then}}}
\newcommand{\Var}{\mathop{\kw{var}}}
\newcommand{\While}{\mathop{\kw{while}}}

\newcommand{\UndefinedValue}{\lightning}
\newcommand{\Cond}[3]{#1 \mathrel{?} #2 : #3}
\newcommand{\Cand}{\mathbin{\mathbf{cand}}}
\newcommand{\Cor}{\mathbin{\mathbf{cor}}}

\newcommand{\lBrace}{\{\mkern -4.5mu{|}}
\newcommand{\rBrace}{|\mkern -4.5mu{\}}}
\newcommand{\Pre}[1]{\lBrace#1\rBrace}

\newcommand{\Fin}[1]{#1^{\star}}
\newcommand{\Finrel}[1]{#1^{*}}
\newcommand{\Om}[1]{#1^{\omega}}

\newcommand{\atomicrel}[1]{\langle#1\rangle}

\newcommand{\guar}[1]{\Guarantee #1}
\newcommand{\Idle}{\kw{idle}}
\newcommand{\opt}[1]{\mathop{\kw{opt}}#1}
\newcommand{\Post}[1]{\Spec{}{}{#1}}
\newcommand{\rely}[1]{\Rely #1}
\makeatletter
\def\Spec{\@ifnextchar*{\@Spec}{\@@Spec}}
\def\@Spec*#1#2#3{\ifx\@empty#1\else#1:\fi
   \llparenthesis{#2}\ifx\@empty#2\else,~\fi#3\rrparenthesis}
\def\@@Spec#1#2#3{\ifx\@empty#1\else
   \begin{array}{@{}l@{}}#1\end{array}:\fi%
   \llparenthesis{\begin{array}{@{}l@{}}#2\end{array}}\ifx\@empty#2\else~,~~\fi
   \begin{array}{@{}l@{}}#3\end{array}\rrparenthesis}
\makeatother
\newcommand{\Term}{\kw{term}}

\newenvironment{RelatedWork}{\paragraph{Related work.}}{\hfill$\Box$}

\newcommand{\opsemantics}[1]{\mathbin{\widehat{#1}}}
\newcommand{\EqEval}[2]{\Set{#1 = #2}}

\newcommand{\LEEval}[2]{\Set{#1 \leq #2}}

\makeatletter
\def\Set{\@ifnextchar*{\@Set}{\@@Set}}
\def\@Set*#1{{\color{blue}\left\llcorner\begin{array}{l}#1\end{array}\right\lrcorner}}
\def\@@Set#1{{\color{blue}\llcorner#1\lrcorner}}
\def\Rel{\@ifnextchar*{\@Rel}{\@@Rel}}
\def\@Rel*#1{{\color{ACMPurple}\left\ulcorner\begin{array}[t]{l}#1\end{array}\right\urcorner}}
\def\@@Rel#1{{\color{ACMPurple}\ulcorner#1\urcorner}} 
\def\RelA{\@ifnextchar*{\@RelA}{\@@RelA}}
\def\@RelA*#1{{\color{ACMPurple}\left\lceil\BB\left\lceil\begin{array}{l}#1\end{array}\right\rceil\BB\right\rceil}}
\def\@@RelA#1{{\color{ACMPurple}\lceil\BB\lceil#1\rceil\BB\rceil}}
\def\Ratomicrel{\@ifnextchar*{\@Ratomicrel}{\@@Ratomicrel}}
\def\@Ratomicrel*#1{\left\langle\Rel*{#1}\right\rangle}
\def\@@Ratomicrel#1{\atomicrel{\Rel{#1}}}
\makeatother

\newcommand{\Rrely}[1]{\rely{{\Rel{#1}}}}
\newcommand{\Rguar}[1]{\guar{{\Rel{#1}}}}
\newcommand{\RPost}[1]{\Post{\Rel{#1}}}

\newcommand{\defs}{\mathrel{\widehat=}}
\renewcommand{\implies}{\mathrel{\Rightarrow}}
\renewcommand{\iff}{\mathrel{\equiv}}
\newcommand{\bool}{\mathbb{B}}
\newcommand{\True}{\mathsf{true}}
\newcommand{\False}{\mathsf{false}}
\newcommand{\dom}{\mathop{\kw{dom}}}
\makeatletter
\def\comp@sym{\raise 0.6ex\hbox{\small\oalign{\hfil%
        $\scriptscriptstyle\mathrm{o}$\hfil%
        \cr\hfil$\scriptscriptstyle\mathrm{9}$\hfil}}}
\newcommand{\semi}{\mathrel{\comp@sym}}
\makeatother
\newcommand{\compl}[1]{\overline{#1}}
\newcommand{\id}[1]{{\textstyle\mathsf{id}}_{#1}}
\newcommand{\universalrel}{\mathsf{univ}}
\newcommand{\dres}{\mathbin{\vartriangleleft}}
\newcommand{\rres}{\mathbin{\vartriangleright}}
\newcommand{\inter}{\mathbin{\cap}}

\newcommand{\union}{\mathbin{\cup}}

\newcommand{\spot}{\mathrel{.}}

\newcommand{\Triple}[3]{\{#1\}\; #2\; \{#3\}}
\newcommand{\TripleV}[3]{\begin{array}{l} \{#1\} \\ #2 \\ \{#3\} \end{array}}
\newcommand{\Establish}[4]{\Triple{#1}{\rely{#2} \together #3}{#4}}
\newcommand{\EstablishV}[4]{\TripleV{#1}{\rely{#2} \together #3}{#4}}
\newcommand{\EstablishExpr}[5]{\Establish{#1}{#2}{\Expr{#3}{#4}}{#5}}
\newcommand{\EstablishLExpr}[5]{\Establish{#1}{#2}{\LExpr{#3}{#4}}{#5}}
\newcommand{\EstablishExprV}[5]{\EstablishV{#1}{#2}{\Expr{#3}{#4}}{#5}}
\makeatletter
\newcommand{\ChainRel}[1]{\crcr \noalign{\penalty\interdisplaylinepenalty}
  \hspace*{-1em}#1 &
  \@ifnextchar*{\@ChainRelCommment}{}}
\newcommand{\Why}[1]{\mbox{{\color{ACMDarkBlue}\hspace*{0.5em}#1}}}
\def\@ChainRelCommment*[#1]{\Why{#1}
  \crcr & 
  }
\newcommand{\StartRef}[1]{\hspace*{-1.5em} \ref{#1}) \refsto
  \@ifnextchar[{\@StartRefCommment}{}}
\def\@StartRefCommment[#1]{\mbox{#1}
  \crcr \noalign{\penalty\interdisplaylinepenalty}}
\makeatother
\newcommand{\Implies}{\ChainRel{\implies}}

\newcommand{\entails}{\Rrightarrow}
\newcommand{\Entails}{\ChainRel{\entails}}
\newcommand{\Equiv}{\ChainRel{\equiv}}
\newcommand{\Refsto}{\ChainRel{\refsto}}
\newcommand{\Equals}{\ChainRel{=}}

\makeatletter
\def\@setmcodes#1#2#3{{\count0=#1 \count1=#3
  \loop \global\mathcode\count0=\count1 \ifnum \count0<#2
  \advance\count0 by1 \advance\count1 by1 \repeat}}

\DeclareSymbolFont{italic}{OT1}{\rmdefault}{m}{it}
\let\mathit\undefined
\DeclareSymbolFontAlphabet{\mathit}{italic}
\edef\@tempa{\hexnumber@\symitalic}
\@setmcodes{`A}{`Z}{"7\@tempa41}
\@setmcodes{`a}{`z}{"7\@tempa61}
\makeatother

\newcommand{\InfRule}[3]
{
    { \boxed{#1} 
      \frac {\strut\displaystyle \begin{array}{c}#2 \end{array}} 
            {\strut\displaystyle \begin{array}{c}#3 \end{array}} 
      \hfil }
}

\usepackage[colorinlistoftodos,textwidth=2cm]{todonotes}

\definecolor{CJ}{rgb}{1,1,0.9}
\definecolor{IH}{rgb}{1,0.9,1}
\definecolor{LM}{rgb}{0.9,1,1}

\newcounter{hours}
\newcounter{minutes}
\newcommand{\printtime}{%
  \ifthenelse{\value{hours}<10}{0}{}\thehours:%
  \ifthenelse{\value{minutes}<10}{0}{}\theminutes}

\newbox{\MyDate}
\savebox{\MyDate}{\draftonly{ \textsf{\footnotesize(\today\ \printtime)}}}

\acmJournal{FAC}

\title[Reasoning about expression evaluation under interference\usebox{\MyDate}]{Reasoning about expression evaluation under interference}

\author[I. J. Hayes]{Ian J. Hayes}
\orcid{0000-0003-3649-392X}
\affiliation{
	\department{School of Electrical Engineering and Computer Science}
	\institution{The University of Queensland}
	\city{Brisbane} 
	\state{Queensland}
	\country{Australia}}
\email{Ian.Hayes@uq.edu.au}

\author[C. B. Jones]{Cliff B. Jones}
\orcid{0000-0002-0038-6623}
\affiliation{
	\department{School of Computing}
	\institution{Newcastle University}
	\city{Newcastle upon Tyne}
	\country{U.K.}}
\email{cliff.jones@ncl.ac.uk}

\author[L. A. Meinicke\usebox{\MyDate}]{Larissa A. Meinicke}
\orcid{0000-0002-5272-820X}
\affiliation{
	\department{School of Electrical Engineering and Computer Science}
	\institution{The University of Queensland}
	\city{Brisbane} 
	\state{Queensland}
	\country{Australia}}
\email{L.Meinicke@uq.edu.au}

\ccsdesc[500]{Theory of computation~Program verification}
\ccsdesc[500]{Theory of computation~Program specifications}
\ccsdesc[500]{Theory of computation~Assertions}
\ccsdesc[500]{Theory of computation~Pre- and post-conditions}
\ccsdesc[500]{Theory of computation~Parallel computing models}

\keywords{shared-memory concurrency, rely-guarantee reasoning, formal methods, specification, program verification, program algebra} 

\begin{abstract}
Hoare-style inference rules for program constructs permit the copying of expressions and tests from program text into logical contexts.
It is known that this requires care even for sequential programs
but much more serious issues arise with concurrent programs because of potential interference to the values of variables.
The ``rely-guarantee'' approach tackles the challenge of recording acceptable interference
and offers a way to provide safe inference rules for concurrent constructs.
This paper shows how the algebraic presentation of rely-guarantee ideas
can clarify and formalise the conditions for safely re-using expressions and tests from program text in logical contexts for reasoning about concurrent programs;
crucially this extends to handling expressions that reference more than one shared variable.
A non-trivial example related to the Fischer-Galler forest representation of equivalence relations is treated.
\end{abstract}

\begin{document}

\ifarxiv
\else
\let\origthepage=\thepage
\makeatletter
\renewcommand{\thepage}{\@arabic\c@page-R}
\makeatother
\clearpage
\let\thepage=\origthepage
\setcounter{page}{1}
\setcounter{section}{0}
\write128{Start of main material: \the\ReadonlyShipoutCounter.}
\fi

\maketitle

\section{Introduction}\labelsect{introduction}

For sequential programming constructs,
inference rules in Hoare style~\cite{Hoare69a} are well known but must be used with care.
For example rules for compound commands such as:
\begin{align*}
\InfRule
{\If}
{{\Triple{p \land b}{c_1}{q}}\\
\Triple{p \land \neg b}{c_2}{q}}
{\Triple{p}{\If b \Then c_1 \Else c_2 \Fi}{q}}
\hspace{2em}
\InfRule
{\While}
{{\Triple{p \land b}{c}{p}}}
{\Triple{p}{\While b \Do c \Od}{p \land \neg b}}
\end{align*}
employ copying of boolean expressions ($b$) from program texts into pre conditions.
Similarly,
another inference rule permits the copying of general expressions from the right-hand side of assignment commands into logical contexts:
\begin{align*}
\InfRule
{\Assign}
{}
{\Triple{p[v \backslash e]}{\Assignment{v}{e}}{p}}
\end{align*}
It is known that this assignment axiom requires care even for sequential programs
(e.g.~where indexed array variables are referenced using subscripts)
see \cite[\S5.2]{AptOlderog-19}.

Additional issues must be addressed for concurrent programs because of potential interference to the values of variables mentioned in expressions.
For example, it should be clear that:
\begin{align*}
  \If v \neq 0 \Then \Assignment{u}{w \div v} \Else \Assignment{u}{0} \Fi \parallel \Assignment{v}{0}
\end{align*}
should not carry the truth of the test condition to a pre condition for the then clause 
because interleavings with
$\Assignment{v}{0}$
can negate the outcome and give rise to a division by zero.

Furthermore, with an assignment to a local variable of the value of a shared variable, $\Assignment{w}{v}$,
it  is not safe to use the assignment rule above to conclude that $w = v$ after this assignment if another thread can change $v$.
The problem of interference becomes more convoluted if there are multiple shared variables contained in expressions.
Many papers on concurrency assume that expressions in tests and
even whole assignment commands are executed atomically
(see the historical discussion in~\cite{Jones-23-MaM}). 
With realistic programming languages and compilers,
this assumption is unwarranted
because assignment commands whose right-hand sides contain expressions with multiple variable references 
will be translated into sequences of machine instructions that access variables separately.
For example, with $v$ as an integer:
$\Assignment{v}{2 * v}$ and
$\Assignment{v}{v + v}$
might be considered to be equivalent in a sequential program but the latter might yield an odd value if a concurrent thread can change the value of $v$ between the two accesses;
only the former satisfies the post condition $even(v')$.
It is obvious that there are many examples of such complications with concurrency
(e.g.~$FINDP$ from \cite{Owicki75} uses test expressions that contain references to shared variables
and the loop test expressions in the example in \refsect{equiv-example} refer to multiple shared variables).

In~\cite{Hoare71a},
Hoare made the important step of using his axioms as decomposition rules to facilitate the formal development of programs.
It is useful to think of this as reading the inference rules upside down:
from the conclusion to what sub-goals need to be achieved.

The crucial property of compositionality was straightforward for sequential program constructs
but is more elusive for concurrency
(see~\cite{DeRoever01}).
The ``rely-guarantee'' approach~\cite{Jones81d,Jones83a,Jones83b} tackles the issue of recording acceptable interference
and offers a way to provide safe inference rules that also support a notion of compositional development.
The basic rely-guarantee idea is simple:
if one is to reason about (portions of) programs that are to function in the presence of interference,
there must be assumptions about the extent of that interference.
As indicated in~\reffig{rely-guar}, 
rely and guarantee conditions are relations between states
(just as post conditions are).
The decision to record acceptable and inflicted interference as relations
limits expressivity but has proved adequate for useful examples of developing programs  that use concurrency ---
see~\cite{HayesJones18}.

\begin{figure}[h]
\begin{center}
\input{rely-guar}
\end{center}
\Description{The figure shows a trace consisting of program and environment transitions,
explicitly seven transitions environment, program, environment, environment, program, program, and environment,
with each environment transition satisfying the rely condition $r$ 
and each program transition satisfying the guarantee condition $g$.
The states of the trace are $\sigma_0$ through to $\sigma_7$.
The initial state $\sigma_0$ satisfies the precondition $p$
and the postcondition relation $q$ is satisfied between the initial state $\sigma_0$ and the final state $\sigma_7$.
The trace is terminating, as indicated by a checkmark at the end.}
\caption{A trace satisfying a specification with precondition $p$, postcondition $q$, rely condition $r$ and guarantee condition $g$.
If the initial state, $\sigma_0$, is in $p$ and all environment transitions ($\estepd$) satisfy $r$,
then all program transitions ($\pstepd$) must satisfy $g$ and 
the initial state $\sigma_0$ must be related to the final state $\sigma_7$ by the postcondition $q$, also a relation between states.
The $\checkmark$ at the right indicates a terminating trace.
}\labelfig{rely-guar}
\end{figure}

Inference rules can be provided that can be used to verify the decomposition of a specified component
into compositions of sub-components.
To make it easier to relate rely-guarantee rules with the Hoare-triples above,
it is possible%
\footnote{In practical examples, these conditions are often too long for this five-tuple form 
and realistic specifications can be presented in a two dimensional layout with keywords.
Furthermore the algebraic presentation in the remainder of this paper offers a uniform way of combining the clauses of a specification.}
to write the pre and rely conditions in braces before the command and the guarantee and post conditions afterwards
$\Triple{p,r}{c}{g,q}$.
So, for example,
a rule for conditional commands such as:
\begin{align}\labelprop{parallel-if}
\InfRule
{\parallel$-$\If}
{{\Triple{p \land b, r}{c_1}{g, q}}\\
\Triple{p \land \neg b, r}{c_2}{g, q}}
{\Triple{p, r}{\If b \Then c_1 \Else c_2 \Fi}{g, q}}
\end{align}

\noindent
partially resolves the issue identified above with the Hoare rule for conditionals because
rely conditions express interference that can occur anywhere including before and after the specified commands execute
(see~\reffig{rely-guar}).
Thus the fact that the test $b$ is true when evaluated in the if command does not establish that it will still hold at the beginning of execution of $c_1$
({\em mutatis mutandis} for $c_2$).

But copying program text into a precondition ignores the fact that it might not be evaluated in a single state (let alone in the initial state which is assumed to satisfy $p$).
For example, in reasoning about,
$\If x = x \Then c_1 \Else c_2 \Fi$,
copying $x \neq x$ as a precondition for $c_2$ is not to be treated as equivalent to $\False$ because the two accesses to $x$ might be made in different states.%
\footnote{Care is required even in sequential cases and \cite{Jones90a} has side conditions on the relationship between expressions in logic
and their use in program constructs.
In neither sequential nor concurrent contexts is a developer likely to write $x = x$ as a test in say a conditional 
but  it is important to note that negating this tautology would yield $\False$ as a precondition for the else clause and allow any nonsense implementation;
an execution which experienced interference between the two accesses of the value of variable $x$ could however execute the else branch.}
The problem of multiple references to shared variables is exactly what this paper is intended to resolve.

It is important to recognise that the motivation for the rely-guarantee approach is to support compositional development 
(rather than post-facto justification) 
of programs.
Thus, 
because the rely condition does limit the interference, 
it is normally straightforward to reason about properties that can be inferred prior to the execution of each branch. 
So, for example, 
$\If v \leq w \Then c_1 \Else c_2 \Fi$,
might,
if $v$ and/or $w$ are shared variables,
leave $v > w$ at the start of execution of $c_1$;
but, in a sensible decomposition of a specification to use a conditional,
a rely condition $v' \leq v \land w \leq w'$
(where unprimed variable names stand for their value in the pre state and primed variables their value in the post state)
would ensure that inheriting the condition $v \leq w$ as a pre condition for $c_1$ is safe.
Note that under the same rely condition, if $v \leq w$ evaluates to false, 
one cannot assume $v > w$ as a precondition of $c_2$,
in fact the strongest condition one can assume is $\True$.
See Example \ref{ex-v-le-u} for a formal treatment.
This class of issues is handled naturally by rely-guarantee rules because the rely conditions distribute into the sub-commands,
for example, into a sequential composition and into a conditional command 
(see \refprop{rely-distrib-seq} and \reflaw*{intro-conditional} below).

As recognised in~\cite{DeRoever01}, rely-guarantee ideas achieve a notion of compositionality that makes it possible 
to decompose a specified objective into sub-components that are to be executed concurrently.
This paper extends that notion of compositionality to expression evaluation under interference from concurrent threads.
There remains the question of soundness of rules such as $\parallel$-$\If$ \refprop{parallel-if}
and the remainder of this paper shows that this issue can be addressed by straightforward algebraic reasoning
rather than having to prove soundness with respect to an operational semantics
or employ restrictive syntactic side conditions.

\paragraph{Issues with existing approaches.}

In common with most of the material on development approaches to concurrency,
early publications on rely-guarantee ideas 
\cite{Jones81d,Jones83a,Jones83b,Stolen90,stolen1991method,XuRoeverHe97,Dingel02,PrensaNieto03,Wickerson10-TR,DBLP:conf/esop/WickersonDP10,SchellhornTEPR14,Sanan21} 
treat expression evaluation in conditionals and loops as atomic
and assignment commands as a whole are considered to execute atomically.
Clearly this is an unrealistic assumption for current programming language implementations.
To get around this issue, 
many approaches require that each expression evaluation or assignment command 
have at most a single occurrence of a critical reference.
\begin{definitionx}[critical-reference]
An occurrence of a variable is defined to be a \emph{critical reference}:
(a) if it is assigned to in one thread and has an occurrence in another thread, or
(b) if it has an occurrence in an expression in one thread and is assigned to in another \cite{BenAri2005}.
\end{definitionx}
For an assignment command, the single critical reference may either be on the left or right side but not both.
Note that the check for a single critical reference is syntactic.
While this constraint has a long history \cite{Bernstein66},%
\footnote{There are publications going back at least as far as~\cite{OwickiGries76} 
that employ the restriction of limiting any expression to have at most one critical reference
(this rule is sometimes mistakenly attributed to John Reynolds).
Expanding out an expression with multiple critical references to use local temporary variables of course exposes (rather than solves) the problem.}
the validity of the syntactic constraint cannot be proven in the earlier rely-guarantee approaches,
precisely because of the atomicity assumptions in their semantics.
In distinction to the earlier approaches is the clear identification of the issue of atomicity in \cite{CoJo07,Coleman08}
which use a fine-grained operational semantics 
that does not assume expression evaluation or assignments are atomic; 
however, the inference rules developed in that work, although practically useful, are limited.
One contribution of this paper is to provide a precise semantic study of expression evaluation under interference.

To avoid any confusion it should be made clear that the notion of interference is inherent in the concept of state in imperative concurrent programming languages.
This paper uses what are often referred to as ``stack variables'' in examples
but exactly the same issues around interference are present with heap variables.
Thus --despite the many strengths of ``Concurrent Separation Logic'' (CSL) \cite{OHearn07}--
it would not resolve the interference issues addressed in this paper.
Our theory abstracts state and thus allows state to contain either stack variables or heap variables or a combination of both.
The theory deals with sets of (abstract) states. 
(The current authors are also fully aware of the interesting combinations of CSL with rely-guarantee ideas:
RGSep \cite{marriage-RGSep,Vafeiadis07}
and 
SAGL \cite{FengSAGL-07}.
In fact the idea in~\cite{JonesYatapanage-15} of viewing separation as an abstraction
that must be achieved when reifying stack variables to heap representations might be more relevant.)

\paragraph{Our approach}

In order to give a fine-grained semantics of expressions,
we represent the evaluation of an expression $e$ to a value $k$ by the command, $\Expr{e}{k}$,
which terminates if $e$ evaluates to $k$ but becomes infeasible otherwise \cite{DaSMfaWSLwC,hayes2021deriving}.
The command $\Expr{e}{k}$ is typically used within a non-deterministic choice over $k$,
in which for any particular evaluation of expression $e$ only one of the choices can be feasible.
Treating an expression evaluation as a command%
\footnote{This should not be confused with programming languages, like C,
in which all commands are treated as expressions (possibly of type void).
Our expressions preclude side effects.}
in our rely-guarantee theory allows one to explicitly treat the effects of interference on expressions.

In Hoare logic the triple $\Triple{p}{c}{p_1}$ states that 
``a command $c$ establishes the postcondition $p_1$ from initial states satisfying $p$''.
Of course such a triple can be invalidated by interference from concurrent threads
and this is where one can make use of a rely condition to bound the allowable interference.
This allows one to generalise establishing a postcondition to 
``a command $c$ \emph{establishes} postcondition $p_1$ from initial states satisfying $p$ 
\emph{under interference satisfying the rely condition $r$}'', 
which can be written as the Hoare triple,
\begin{align}\labelprop{establishes}
  \Establish{p}{r}{c}{p_1} 
\end{align}
where the command, $\rely{r}$, and the operator $\together$ are defined in \refsect{language}.
Because an expression evaluation is a command,
we can apply this generalised concept to state that $\EstablishExpr{p}{r}{e}{k}{p_1}$,
that is, the expression evaluation $\Expr{e}{k}$ establishes postcondition $p_1$ from initial states satisfying $p$
\emph{under interference satisfying rely condition $r$}.%
\footnote{While later we use a relational postcondition, $q$, for a specification command \refdef{spec},
here a single-state postcondition, $p_1$, as used in Hoare logic is sufficient 
because expression evaluation is side-effect free, i.e.\ it does not change the state.}
It turns out that this is just what is required to reason about boolean expressions in the guards of conditionals and while loops
as well as expressions in assignment commands.

\paragraph{Overview}

This paper shows how the algebraic presentation of rely-guarantee ideas~\cite{HayesJones18,HJM-23}
can clarify and formalise the necessary conditions to reason about arbitrary expressions.
The approach,
\begin{itemize}
\item
makes use of a fine-grained theory (see \refsect{language}) without the assumption that expression evaluation and assignment commands are atomic
--- see \refsect{expressions},
\item
allows the development of \emph{compositional} rules for reasoning about expression evaluation 
that do not require the single critical reference syntactic constraints
--- see \refsect{complex-expressions},
\item
allows the development of refinement rules for introducing conditionals, assignment commands and while loops
--- see \refsect{commands},
\item
allows application to programs without syntactic restrictions on expressions and assignment commands
-- see \refsect{equiv-example} for an application to a non-trivial example,
and
\item
the concurrency theory is formalised within the Isabelle/HOL theorem prover \cite{IsabelleHOL}
and all the lemmas presented in this paper have been proven in that theory.
\end{itemize}

\section{Wide-spectrum language}\labelsect{language}

Our wide-spectrum language \cite{CIP} is modeled on that of the sequential refinement calculus \cite{BackWright98,Morgan94}
extended with a parallel operation,
which requires a richer trace-based semantics \cite{DaSMfaWSLwC}.

\paragraph{Naming and syntactic precedence conventions.}

We use 
$\Sigma$ for the set of program states,
$\sigma$ for program states (i.e.\ $\sigma \in \Sigma$),
$c$ and $d$ for commands;
$p$ for sets of program states;
$g$, $q$ and $r$ for binary relations between program states;
$u$, $v$ and $w$ for program variables;
$e$ for expressions,
$b$ for boolean expressions, where $\bool$ is the set of booleans,
and 
$k$ and $\kappa$ for values.
Subscripted versions of the above names follow the same convention.
Unary operations and function application have higher precedence than binary operations.
For binary operations, non-deterministic choice ($\nondet$) has the lowest precedence, 
and sequential composition ($\Seq$) has the highest precedence.
We use parentheses to resolve all other syntactic ambiguities.

\subsection{Command lattice and primitive operations}\labelsect{lattice}

Our language consists of a complete lattice of commands
with partial order, $c \refsto d$, meaning command $c$ is refined (or implemented) by command $d$,
so that non-deterministic choice ($c \nondet d$) is the lattice join 
and strong conjunction ($c \meet d$) is the lattice meet \cite{AFfGRGRACP,Concurrent_Ref_Alg-AFP,FM2016atomicSteps,FMJournalAtomicSteps}.
The lattice is complete so that, for a set of commands $C$, 
non-deterministic choice, $\Nondet C$, and strong conjunction, $\Meet C$, are defined 
as the least upper bound and greatest lower bound, respectively.
The lattice operators $\nondet$ and $\meet$, and their counterparts over sets of commands $\Nondet$ and $\Meet$,
are considered part of our wide spectrum language.

The semantic model for the lattice of commands consists of sets of Aczel traces \cite{Aczel83,DeRoever01,DaSMfaWSLwC}.
An Aczel trace consists of a sequence of state-to-state transitions 
each of which is labeled as either a program ($\pstepd$) or environment ($\estepd$) transition, as in \reffig{rely-guar}.
A trace may be either terminating, aborting or incomplete.
Sets of traces corresponding to commands are 
prefix closed (i.e.\ they contain all incomplete prefix traces of a trace)
and abort closed (i.e.\ they contain all possible extensions of an aborting trace).
The language includes the following three binary operators:
\begin{description}
\item[$c \Seq d$,]
sequential composition, that is associative, has the null command, $\Nil$, as its neutral element,
and distributes arbitrary non-deterministic choices in its left argument \refax{Nondet-distrib-seq-right} 
and non-empty non-deterministic choices in its right argument \refax{Nondet-distrib-seq-left}.
A trace of $c \Seq d$ consists of the gluing concatenation of a terminating trace of $c$ and a trace of $d$,
or an incomplete or aborting trace of $c$.
A gluing concatenation of a terminating trace $tr_1$ with a trace $tr_2$ 
requires that the final state of $tr_1$ equals the initial state of $tr_2$ 
and these two states become a single state in the concatenation.
\item[$c \parallel d$,]
parallel composition, that is associative, commutative, 
and distributes non-empty non-deterministic choices in both its arguments \refax{Nondet-distrib-par} 
because $\parallel$ is commutative.
Because each trace records \emph{both} the behaviour of the program \emph{and} its environment, parallel composition is synchronous, and 
a trace of $c \parallel d$ consists of a matching of a trace of $c$ and a trace of $d$,
where a program transition $\ptranssp$ of $c$ matches an environment transition $\etranssp$ of $d$ to give $\ptranssp$ for $c \parallel d$ 
(or vice versa)
and an environment transition $\etranssp$ for both $c$ and $d$ matches to give the same environment transition for $c \parallel d$.
Note that a program transition of $c$ cannot match a program transition of $d$,
leading to the interleaving of their program transitions.
If either $c$ or $d$ aborts, so does $c \parallel d$.
Termination of $c \parallel d$ corresponds to termination of both $c$ and $d$.
\item[$c \together d$,]
weak conjunction, that is associative, commutative, idempotent, 
and distributes non-empty non-deterministic choices in both its arguments \refax{Nondet-distrib-conj} 
because $\together$ is commutative.
A trace of $c \together d$ is either a trace of both $c$ and $d$, 
or it is an aborting trace of $c$ that equals an incomplete trace of $d$ up to the point of abort of $c$ (or vice versa).
Weak conjunction ($c \together d$) behaves like strong conjunction ($c \meet d$) unless either operand aborts, 
in which case the weak conjunction aborts.
Termination of $c \together d$ corresponds to termination of both $c$ and $d$.
\end{description}
\begin{align}
  \Nondet_{c \in C} (c \Seq d) & = (\Nondet_{c \in C} c) \Seq d \labelax{Nondet-distrib-seq-right} \\
  \Nondet_{c \in C} (d \Seq c) & = d \Seq (\Nondet_{c \in C} c)  & \mbox{if } C \neq \emptyset \labelax{Nondet-distrib-seq-left} \\
  \Nondet_{c \in C} (c \parallel d) & = (\Nondet_{c \in C} c) \parallel d  & \mbox{if } C \neq \emptyset \labelax{Nondet-distrib-par} \\
  \Nondet_{c \in C} (c \together d) & = (\Nondet_{c \in C} c) \together d  & \mbox{if } C \neq \emptyset \labelax{Nondet-distrib-conj} 
\end{align}

\subsection{Abstraction of program variables, values and states}\labelsect{state}

Our theory treats values, left values (variables and indexed array elements) and program states abstractly.
For values ($\in Value$) the only constraint is that $Value$ contains the distinct boolean values $\True$ and $\False$
(so that we can handle boolean guards in conditionals and loops).
The left values (l-values for short) of our language consist of base variable names ($v \in string$) and 
indexed array elements ($\LEx{Indexed\,A\,\Ex{i}}$), where $A$ is an l-value and $i$ is an index value.
\begin{align}
  LValue \defs Base\,string \mid Indexed\,LValue\,Value  \labeldef{LValue}
\end{align}
Where it does not cause confusion, we abbreviate ``$\LEx{Base\,v}$'' to ``$\Base{v}$'', 
and ``$\LEx{Indexed\,A\,\Ex{i}}$'' to ``$\Indexed{A}{i}$''.
As a syntactic aid to comprehension, expressions are displayed in a \Ex{blue} font and l-value expressions in a \LEx{purple} font.
L-values could be extended to handle a heap but we do not consider that in this paper.%
\footnote{Because a heap corresponds to an anonymous array with indices being pointer values,
the theory for handling arrays could be used to handle a heap.}
The set of possible program states, $\Sigma$, is a total mapping from l-values to values:
\begin{align}
  \Sigma \defs LValue \fun Value \labeldef{state}
\end{align}
where the inclusion of an undefined value, $\UndefinedValue$, in
$Value$, for example, can allow us to represent the value of
undefined l-values in the state, e.g. to represent the value
of an undeclared variable, or the value of an out-of-bounds array access.

For our theory it is sufficient to view the state through a lens 
\cite{BackWright98,2020FosterSimonUnifyingSemanticFoundations,2007FosterNathanCombinatorsForBidirectional} 
formed by two functions,
$get\,lv\,\sigma$, that returns the value of l-value $lv$ in state $\sigma$ and,
$set\,lv\,k\,\sigma$, that returns a state that differs from $\sigma$ only at $lv$ which has the value $k$.
They have the following types,
\begin{align}
  get & : LValue \fun \Sigma \fun Value \labeldef{get} \\
  set & : LValue \fun Value \fun \Sigma \fun \Sigma \labeldef{set}
\end{align}
and satisfy the standard lens axioms 
\cite{BackWright98,2020FosterSimonUnifyingSemanticFoundations,2007FosterNathanCombinatorsForBidirectional}.
The $get$ and $set$ operations atomically access the state
and hence we are assuming that base variable accesses are atomic and
accesses to an indexed array element are atomic.
We assume l-values used within program code are fully indexed,
that is, we do not allow access to whole arrays or array slices in program code.

\subsection{Primitive commands}\labelsect{primitives}

A command is infeasible in a state, $\sigma$, if it cannot make any transition from $\sigma$
and it can neither terminate nor abort in state $\sigma$.
The least command in our lattice, $\Magic$, is everywhere infeasible.
The language includes four primitive commands:
\begin{description}
\item[$\cgd{p}$]
is an instantaneous test that the current state $\sigma$ is in the set of states $p$:
if $\sigma \in p$, $\cgd{p}$ terminates, otherwise it is infeasible -- 
note that there is no transition taken by an instantaneous test 
(in the trace semantics, its trace is tagged as terminated);
\item[$\cpstep{r}$]
is an atomic program command 
that may perform a program transition $\ptranssp$ if $(\sigma,\sigma') \in r$ and then terminate,
otherwise it is infeasible if $\sigma \not\in \dom r$;
\item[$\cestep{r}$]
is an atomic environment command 
that may perform an environment transition $\etranssp$ if $(\sigma,\sigma') \in r$ and then terminate,
otherwise it is infeasible if $\sigma \not\in \dom r$; and
\item[$\Abort$~]
is Dijkstra's abort command \cite{Dijkstra75,Dijkstra76} that can do any behaviour whatsoever.
It is the greatest command.
The aborting behaviour is irrecoverable, that is, $\Abort \Seq c = \Abort$ for any command $c$.
\end{description}
For tests, $\cgd{p_1} \refsto \cgd{p_2}$ if $p_1 \supseteq p_2$,
and both $\cpstep{r_1} \refsto \cpstep{r_2}$ and $\cestep{r_1} \refsto \cestep{r_2}$ hold if $r_1 \supseteq r_2$.
Tests/atomic commands are everywhere infeasible for the empty set/relation:
$\cgd{\emptyset} = \cpstep{\emptyset} = \cestep{\emptyset} = \Magic$, the least command.
A command is considered \emph{atomic} if it is of the form, $\cpstep{g} \nondet \cestep{r}$, for some relations $g$ and $r$,
that is, it can only make a single transition, 
which may be a program ($\pstepd$) transition in $g$ or an environment ($\estepd$) transition in $r$.

The command $\Nil$ is the test that always succeeds \refdef{nil}.
The assert command, $\Pre{p}$, aborts if the current state is not in $p$, 
otherwise it terminates immediately \refdef{pre}.
Property \refprop{pre-alt} gives an alternative form for an assert command.
Note that $\compl{p} = \Sigma - p$ and $\Pre{\emptyset} = \Abort$ and $\Pre{\Sigma} = \Nil$.
The command $\cstep{r}$ allows either a program or an environment transition, provided it satisfies $r$ \refdef{cstep}.
The abbreviations $\cpstepd$ and $\cestepd$ allow any program \refdef{cpstepd} or environment \refdef{cestepd} transition, respectively,
and $\cstepd$ allows any transition, program or environment \refdef{cstepd},
where $\universalrel$ is the universal relation between program states \refdef{univ}.
Note the bold fonts for $\Nil$, $\cpstepd$, $\cestepd$ and $\cstepd$.
\\
\begin{minipage}{0.5\textwidth}
\begin{align}
  \Nil & \defs \cgd{\Sigma} \labeldef{nil} \\
  \Pre{p} & \defs \Nil \nondet \cgd{\compl{p}} \Seq \Abort \labeldef{pre} \\
  \Pre{p} & = \cgd{p} \nondet \cgd{\compl{p}} \Seq \Abort \labelprop{pre-alt} \\
  \cstep{r} & \defs \cpstep{r} \nondet \cestep{r} \labeldef{cstep} 
\end{align}
\end{minipage}%
\begin{minipage}{0.5\textwidth}
\begin{align}
  \universalrel & \defs \Sigma \times \Sigma \labeldef{univ} \\
  \cpstepd & \defs \cpstep{\universalrel} \labeldef{cpstepd} \\  
  \cestepd & \defs \cestep{\universalrel} \labeldef{cestepd} \\  
  \cstepd & \defs \cstep{\universalrel} \labeldef{cstepd}  
\end{align}
\end{minipage}
\\[1ex]
Given a set of states $p$ and a relation $r$, 
$p \dres r$ is the relation $r$ restricted so that its domain elements are within $p$,
and
$r \rres p$ is the relation $r$ with its range elements restricted to be in $p$.
\begin{align}
  p \dres r & \defs \{(\sigma,\sigma') \in r \spot \sigma \in p\} \labeldef{dres} \\
  r \rres p & \defs \{(\sigma,\sigma') \in r \spot \sigma' \in p\} \labeldef{rres}
\end{align}
The basic commands satisfy the following refinement properties.
\\\begin{minipage}[t]{0.5\textwidth}
\begin{align}
  \Pre{p} \Seq \cgd{p_1} & \refsto \cgd{p_2}  &\mbox{if } p_1 \supseteq p \inter p_2 \labelprop{cgd-ref} \\
  \Pre{p} \Seq \cpstep{r_1} & \refsto \cpstep{r_2} & \mbox{if } r_1 \supseteq p \dres r_2 \labelprop{cpstep-ref} \\
  \Pre{p} \Seq \cestep{r_1} & \refsto \cestep{r_2}  &\mbox{if } r_1 \supseteq p \dres r_2 \labelprop{cestep-ref} \\
  \Pre{p_1} & \refsto \Pre{p_2}  & \mbox{if } p_1 \subseteq p_2 \labelprop{assert-weaken} \\
  \Pre{p_1} \Seq \cgd{p_2} & \refsto \cgd{p_2} \Seq \Pre{p_3} & \mbox{if } p_1 \inter p_2 \subseteq p_3 \labelprop{assert-plus-test} 
\end{align}
\end{minipage}%
\begin{minipage}[t]{0.5\textwidth}
\begin{align}
  \Pre{p_1} \Seq \Pre{p_2} & = \Pre{p_1 \inter p_2} \labelprop{assert-merge} \\
  \Pre{p} \Seq \cgd{p} & = \Pre{p} \labelprop{assert-test}  \\
  \Pre{p} \Seq c & \refsto c \labelprop{remove-assert} \\
  \Pre{p} \Seq (c \parallel d) & = (\Pre{p} \Seq c) \parallel (\Pre{p} \Seq d) \labelprop{assert-distrib-par}\\
  (c \Seq \Pre{p}) \parallel d & \refsto (c \parallel d) \Seq \Pre{p} \labelprop{par-assert-after}
\end{align}
\end{minipage} \\

\subsection{Derived commands}\labelsect{derived}

Finite iteration \refdef{finite-iter}, $\Fin{c}$, and possibly infinite iteration \refdef{iter}, $\Om{c}$, 
are defined as the least ($\mu$) and greatest ($\nu$) fixed points, respectively, of the monotone function $(\lambda y \spot \Nil \nondet c \Seq y)$.
A program guarantee command, $\guar{r}$ for relation $r$, requires that every program transition satisfies $r$
but places no constraints on environment transitions \refdef{guar}.
A rely command, $\rely{r}$, assumes environment transitions satisfy $r$;
if any environment transition does not, it aborts \refdef{rely},
in the same way that an assertion $\Pre{p}$ aborts if the initial state is not in $p$.
The notation $\compl{r}$ stands for the complement of the relation $r$.
The command $\Term$ only performs a finite number of program transitions
but does not constrain its environment \refdef{term}.
The postcondition specification command, $\Post{q}$, for $q$ a relation between states,
guarantees to terminate in a final state $\sigma'$ that is related to the initial state $\sigma$ by $q$ \refdef{spec}.
Note that the command $\cgd{\{\sigma\}}$ terminates from state $\sigma$ but is infeasible otherwise.
The relation $\id{}$ is the identity relation \refdef{id}.
The command $\Idle$ can perform only a finite number of stuttering (no change) program transitions
but does not constrain its environment \refdef{idle}.
The command $\opt{r}$ either performs a single program transition satisfying $r$
or it can terminate immediately from states $\sigma$ such that $(\sigma,\sigma) \in r$ \refdef{opt},
that is $r$ is satisfied by not changing the state.
\\[-2ex]
\begin{minipage}[t]{0.4\textwidth}
\begin{align}
  \Fin{c} & \defs \mu y \spot \Nil \nondet c \Seq y \labeldef{finite-iter} \\
  \Om{c} & \defs \nu y \spot \Nil \nondet c \Seq y \labeldef{iter} \\
  \guar{r} & \defs \Om{(\cpstep{r} \nondet \cestepd)}  \labeldef{guar} \\
  \rely{r} & \defs \Om{(\cstepd \nondet \cestep{\compl{r}} \Seq \Abort)}  \labeldef{rely} \\
  \Term & \defs \Fin{\cstepd} \Seq \Om{\cestepd}  \labeldef{term} 
\end{align}
\end{minipage}%
\begin{minipage}[t]{0.6\textwidth}
\begin{align}
  \Spec{}{}{q} &\defs \Nondet_{\sigma} \cgd{\{\sigma\}} \Seq \Term \Seq \cgd{\{\sigma' \spot (\sigma,\sigma') \in q\}}  \labeldef{spec} \\
  \id{} & \defs \{(\sigma,\sigma) \spot \sigma \in \Sigma \} \labeldef{id} \\
  \Idle & \defs \guar{\id{}} \together \Term  \labeldef{idle} \\
  \opt{r} & \defs \cpstep{r} \nondet \cgd{\{\sigma \spot (\sigma,\sigma) \in r\}}  \labeldef{opt}
\end{align}
\end{minipage}
\\[1ex]
Finite iteration satisfies the standard least fixed point induction property \refprop{fin-iter-induct}.
Guarantees distribute over sequential \refprop{guar-distrib-seq} and parallel composition \refprop{guar-distrib-par}
and relies distribute over sequential composition \refprop{rely-distrib-seq} \cite{2024MeinickeHayesDistributiveLaws}.
A rely command may be weakened \refprop{rely-weaken}, with the ultimate weakening being to remove it \refprop{rely-remove}.
Property \refprop{triple-with-rely} gives an alternative definition for a Hoare triple with a rely condition.
\begin{align}
  d & \refsto \Fin{c}  \hspace{14.5em}\mbox{if } d \refsto \Nil \nondet c \Seq d \labelprop{fin-iter-induct} \\
  \guar{g} \together c_1 \Seq c_2 & = (\guar{g} \together c_1) \Seq (\guar{g} \together c_2) \labelprop{guar-distrib-seq} \\
  \guar{g} \together (c_1 \parallel c_2) & = (\guar{g} \together c_1) \parallel (\guar{g} \together c_2) \labelprop{guar-distrib-par} \\
  \rely{r} \together c_1 \Seq c_2 & = (\rely{r} \together c_1) \Seq (\rely{r} \together c_2) \labelprop{rely-distrib-seq} \\
  \rely{r_1} & \refsto \rely{r_2} \hspace{13em}\mbox{if } r_1 \subseteq r_2 \labelprop{rely-weaken} \\
  \rely{r} \together c & \refsto c \labelprop{rely-remove} \\
  (\Triple{p}{\rely{r} \together c}{p_1}) & \iff (\rely{r} \together \Pre{p} \Seq c \refsto \rely{r} \together c \Seq \Pre{p_1}) \labelprop{triple-with-rely}
\end{align}

A specification command may be refined by strengthening its post condition \refprop{spec-strengthen}. 
A specification command may be refined by an optional command \refdef{opt} in the context of a precondition $p$
and guarantee condition $g$ provided the optional command also satisfies the guarantee \refprop{spec-to-opt}.
\begin{align}
  \Spec{}{}{q_1} & \refsto \Spec{}{}{q_2} & \mbox{if } q_1 \supseteq q_2 \labelprop{spec-strengthen} \\
  \guar{g} \together \Pre{p} \Seq \Post{q_1} & \refsto \opt{q_2}  & \mbox{if } p \dres q_2 \subseteq g \inter q_1 \labelprop{spec-to-opt}
\end{align}

A rely may be distributed into a parallel composition 
provided each of the commands in the parallel composition guarantees the rely \cite{2024MeinickeHayesDistributiveLaws}.
\begin{lemmax}[rely-par-distrib-with-guar]
For a relation $r$, and a pair of commands $c_1$ and $c_2$,
if $\guar{r} \together c_1 = c_1$ and $\guar{r} \together c_2 = c_2$, 
then $\rely{r} \together (c_1 \parallel c_2) = (\rely{r} \together c_1) \parallel (\rely{r} \together c_2)$.
\end{lemmax}

A reflexive rely may be distributed into a parallel composition of $\Idle$ commands or expression evaluations
because they do not change the state and hence guarantee any reflexive relation.
\begin{lemmax}[rely-distrib-par-idle]
If $r$ is a reflexive relation, and $c_1$ and $c_2$ are commands that refine $\Idle$, 
\begin{align*}
\rely{r} \together (c_1 \parallel c_2) = (\rely{r} \together c_1)  \parallel (\rely{r} \together c_2) .  
\end{align*}
\end{lemmax}

\subsection{Hoare inference rules in a rely context}\labelsect{hoare-inference}

The inclusion of tests, and subsequently assertions, makes it possible to represent and reason about correctness assertions, such as Hoare triples, in the algebra.

\begin{definitionx}[Hoare-triple]
A command $c$ \emph{establishes} postcondition $p_1$ from initial states satisfying $p$, 
written using the Hoare triple, $\Triple{p}{c}{p_1}$, if and only if,%
\footnote{Note if $c$ does not terminate or aborts, \refprop{triple-form} holds 
because the assertion $\Pre{p_1}$ is not reached,
so this is a weak correctness interpretation of Hoare logic \cite{Wright04}.
For our usages here, $c$ is an expression evaluation, which can neither abort nor fail to terminate.}
\begin{align}
  \Pre{p} \Seq c \refsto c \Seq \Pre{p_1}.  \labelprop{triple-form}
\end{align}
\end{definitionx}
Note the different braces used in the assertion commands and for the pre and post conditions in the Hoare triple.
Because the proofs in this paper make use a a rely context,
our Hoare logic inference rules incorporate a rely condition,
with the exception of the Consequence rule, 
which is already valid if $c$ includes a rely condition.
\begin{lemmax}[hoare-inference]
For commands $c$, $c_1$ and $c_2$, a possibly-empty set of commands $C$, 
a relation $r$, 
and 
sets of states $p_0$, $p_1$, $p_2$ and $p_3$,
\\
\begin{minipage}{0.47\textwidth}
\begin{align}
\InfRule{Consequence}
{p_0 \subseteq p_1 \land p_2 \subseteq p_3 \\
\Triple{p_1}{c}{p_2}}
{\Triple{p_0}{c}{p_3}}
\labelprop{hoare-consequence} \\
\InfRule{Sequence}
{\Triple{p_0}{\rely{r} \together c_1}{p_1} \\
\Triple{p_1}{\rely{r} \together c_2}{p_2}}
{\Triple{p_0}{\rely{r} \together c_1 \Seq c_2}{p_2}}
\labelprop{hoare-sequential} 
\end{align}
\end{minipage}%
\begin{minipage}{0.53\textwidth}
\begin{align}
\InfRule{Choice}
{\forall c \in C \spot \Triple{p_0}{\rely{r} \together c}{p_1}}
{\Triple{p_0}{\rely{r} \together \Nondet C}{p_1}}
\labelprop{hoare-Nondet} \\
\InfRule{Parallel}
{\guar{r} \together c_1 = c_1 \\
\guar{r} \together c_2 = c_2 \\
\Triple{p_0}{\rely{r} \together c_1}{p_1} \\
\Triple{p_0}{\rely{r} \together c_2}{p_2}}
{\Triple{p_0}{\rely{r} \together (c_1 \parallel c_2)}{p_1 \inter p_2}}
\labelprop{hoare-parallel} 
\end{align}
\end{minipage}  
\end{lemmax}

\begin{proof}
From \Definition{Hoare-triple} we have that 
pre-conditions can be strengthened and post-conditions can be weakened \refprop{hoare-consequence}, 
which follows directly from \refprop{assert-weaken}.
A sequential composition satisfies a Hoare triple 
if its first command establishes the pre-condition $p_1$ of the second \refprop{hoare-sequential},
which holds by the sequential application of both assumptions and \Definition{Hoare-triple},
\begin{align*}&
  \Pre{p_0} \Seq (\rely{r} \together c_1 \Seq c_2)
 \Equals*[distribute the rely \refprop{rely-distrib-seq}]
 \Pre{p_0} \Seq (\rely{r} \together c_1) \Seq (\rely{r} \together c_2)
 \Refsto*[apply the first assumption]
 (\rely{r} \together c_1) \Seq \Pre{p_1} \Seq (\rely{r} \together c_2)
 \Refsto*[apply the second assumption]
 (\rely{r} \together c_1)\Seq (\rely{r} \together c_2) \Seq \Pre{p_2} 
 \Equals*[distribute the rely \refprop{rely-distrib-seq} in reverse]
 (\rely{r} \together c_1\Seq c_2) \Seq \Pre{p_2} 
\end{align*}
Property \refprop{hoare-Nondet} is trivially satisfied if $C$ is empty 
because $\Nondet_{c \in \emptyset}{c} = \Magic$ and $\Magic$ is a left-annihilator of sequential composition. 
For non-empty $C$, using \Definition{Hoare-triple} we need to show the following.
\begin{align*}&
  \Pre{p_0} \Seq (\rely{r} \together \Nondet_{c\in C}{c})
 \Equals*[distributing the rely by \refax{Nondet-distrib-conj} and the precondition by \refax{Nondet-distrib-seq-left} assuming non-empty $C$]
  \Nondet_{c\in C}{(\Pre{p_0} \Seq (\rely{r} \together c))}
 \Refsto*[as $\Nondet$ is monotone and assumption $\Triple{p_0}{\rely{r} \together c}{p_1}$ for all $c\in C$]
  \Nondet_{c\in C}{((\rely{r} \together c) \Seq \Pre{p_1})}
 \Equals*[distributing the assertion by \refax{Nondet-distrib-seq-right} and the rely by \refax{Nondet-distrib-conj}]
  (\rely{r} \together (\Nondet_{c\in C}{c})) \Seq \Pre{p_1}
\end{align*}
For \refprop{hoare-parallel} the assumptions that $c_1$ and $c_2$ satisfy a guarantee of $r$ allow one 
to apply \reflem{rely-par-distrib-with-guar} to distribute the rely into the parallel. We have,
\begin{align*}&
  \Pre{p_0} \Seq (\rely{r} \together (c_1 \parallel c_2))
 \Equals*[distribute the rely by \reflem*{rely-par-distrib-with-guar} and distributing the assertion by \refprop{assert-distrib-par}]
  (\Pre{p_0} \Seq (\rely{r} \together c_1)) \parallel (\Pre{p_0} \Seq (\rely{r} \together c_2))
 \Refsto*[assumptions $\Triple{p_0}{\rely{r} \together c_1}{p_1}$ and $\Triple{p_0}{\rely{r} \together c_2}{p_2}$ and $\parallel$ is monotone]
  ((\rely{r} \together c_1) \Seq \Pre{p_1}) \parallel ((\rely{r} \together c_2) \Seq \Pre{p_2})
 \Refsto*[distributing both assertions by \refprop{par-assert-after} twice and merging by \refprop{assert-merge}]
  (\rely{r} \together c_1  \parallel \rely{r} \together c_2) \Seq \Pre{p_1 \inter p_2}
 \Equals*[applying \reflem{rely-par-distrib-with-guar} in reverse]
  (\rely{r} \together (c_1  \parallel c_2)) \Seq \Pre{p_1 \inter p_2}
\qedhere
\end{align*}
\end{proof}

\section{Expressions}\labelsect{expressions}

We distinguish between l-values expressions ($\LEx{lve}$) and expressions ($\Ex{e}$).
The syntax of expressions includes 
constants ($\Const{\kappa}$),
dereferencing of l-value expressions ($\Deref{lve}$),
unary operators ($\Unary{\UnaryOp}{e}$),
and
binary operators ($\Binary{e_1}{\BinaryOp}{e_2}$).
L-value expressions consist of either a variable reference $\Variable{v}$ or 
an indexed array reference $\Array{lve}{e}$,
in which $\LEx{lve}$ is an l-value expression (thus allowing multi-dimensional arrays)
and $\Ex{e}$ is an index expression.

\subsection{Evaluating expressions in a single state}\labelsect{expr-single-state}

Evaluating an expression or l-value expression, $\Ex{e}$, in a (single) state, $\sigma$, is denoted $\Eval{e}{\sigma}$ and 
is defined in (\refdef*{eval-const}--\refdef*{leval-array}),
where for an operator $\odot$, $\opsemantics{\odot}$ is the standard semantics of that operator on values.
For a boolean expression $\Ex{b}$ the notation $\Set{b}$ stands for 
the set of states in which $\Ex{b}$ evaluates to $\True$ \refdef{Set}.
\\[-2ex]\begin{minipage}[t]{0.5\textwidth}
\begin{align}
  \Eval{\Const{\kappa}}{\sigma} & \defs \Ex{\kappa} \labeldef{eval-const} \\
  \Eval{\Deref{lve}}{\sigma} & \defs get\,(\LEval{lve}{\sigma})\,\sigma \labeldef{eval-deref} \\
  \Eval{\Unary{\UnaryOp}{e}}{\sigma} & \defs \opsemantics{\UnaryOp} (\Eval{e}{\sigma}) \labeldef{eval-unary} \\
  \Eval{\Binary{e_1}{\BinaryOp}{e_2}}{\sigma} & \defs (\Eval{e_1}{\sigma}) \opsemantics{\BinaryOp} (\Eval{e_2}{\sigma}) \labeldef{eval-binary} 
\end{align}
\end{minipage}%
\begin{minipage}[t]{0.5\textwidth}
\begin{align}
  \LEval{\Variable{v}}{\sigma} & \defs \Base{v} \labeldef{leval-var} \\
  \LEval{\Array{lve}{e}}{\sigma} & \defs \Indexed{\LEval{lve}{\sigma}}{\Eval{e}{\sigma}} \labeldef{leval-array} \\
  \Set{b} & \defs \{ \sigma \spot \Eval{b}{\sigma} = \True\} \labeldef{Set} 
\end{align}
\end{minipage}\\[1ex]
For expressions $\Ex{e_1}$ and $\Ex{e_2}$,
$\Set{e_1 = e_2} = \{\sigma \spot \Eval{e_1 = e_2}{\sigma} = \True\} = \{\sigma \spot \Eval{e_1}{\sigma} = \Eval{e_2}{\sigma}\}$,
and hence for constants $\Ex{\kappa}$ and $\Ex{k}$,
$\Set{\kappa = k} = \{\sigma \spot \Eval{\kappa}{\sigma} = \Eval{k}{\sigma}\} = \{\sigma \spot \Ex{\kappa = k}\}$,
which equals $\Sigma$ if $\Ex{\kappa = k}$ and the empty set, $\emptyset$, otherwise.
For an l-value expression $\LEx{lve}$, 
$\Set{\Deref{lve} = k} = \{\sigma \spot \Eval{\Deref{lve}}{\sigma} =\Eval{k}{\sigma}\} =$ $ \{\sigma \spot get\,(\LEval{lve}{\sigma})\,\sigma = \Ex{k}\}$.
We also make use of a predicate notation for relations,
in which a dereferenced l-value expression, $\Deref{lve}$, stands for its value in the pre-state and 
a primed dereferenced l-value expression, $\Deref{lve'}$, stands for its value in the post-state,
for example, 
$\Rel{\Deref{lve}' \leq \Deref{lve}} = 
\{ (\sigma,\sigma') \spot \Eval{\Deref{lve}}{\sigma'} \leq \Eval{\Deref{lve}}{\sigma} \} =
\{ (\sigma,\sigma') \spot get\,(\LEval{lve}{\sigma'})\,\sigma' \leq get\,(\LEval{lve}{\sigma})\,\sigma \}$.
Note that to distinguish between sets and relations,
we use lower corners for sets of states and upper corners for relations between states.
We make use of these predicate notations within examples.

\subsection{Evaluating expressions in the presence of interference}\labelsect{expr-command}

While the above semantics \refdef{eval-const}--\refdef{leval-array}
is suitable for a sequential programming language,
once concurrency is added to the language, that semantics is inadequate 
because concurrent threads can be updating variables used within an expression while it is being evaluated.
Hence accesses to different variables within an expression, or even multiple accesses to the same variable,
can take place in different states.

In our semantics, interference from concurrent threads is represented by environment ($\estepd$) transitions,
so our semantics for expression evaluation must allow arbitrary interleaving of environment transitions.
Because our expressions are free of side effects,
calculation steps can be represented by program transitions that do not change the state ($\cpstep{\id{}}$),
commonly called \emph{stuttering} transitions,
and any expression evaluation makes only a finite number of such transitions.
Note that the command, $\Idle$, is used in our semantics for expression evaluation to represent both
finite stuttering program transitions and arbitrary environment transitions.

While in a sequential program the value of an expression is uniquely determined by the initial state in which it is evaluated,
in a concurrent context its value depends on the sequence of states that occur during its evaluation.
Hence for a given initial state, 
an expression may evaluate to different values depending on the interleaving environment transitions 
between its accesses to l-values.
Our semantics needs to allow for such non-determinism.
To handle this we represent the evaluation of an expression $\Ex{e}$ to a value $\Ex{k}$ by the command $\Expr{e}{k}$.
For example, if $\Ex{b}$ is a boolean expression, $\Expr{b}{\True}$ represents evaluating $\Ex{b}$ to $\Ex{\True}$.
The evaluation of an expression $\Ex{e}$ to any possible value can be represented as a non-deterministic choice over $\Ex{k}$, e.g. $\Nondet_{\Ex{k}} \Expr{e}{k}$,
although typically expression evaluation is used to determine control flow  (see \refsect{commands}) and expression evaluation is used in the context of a non-deterministic choice of the form:
\begin{align}
  \Nondet_{\Ex{k}} \Expr{e}{k} ; c\,\Ex{k} 
\labeldef{expr-usage}
\end{align}
where for each $\Ex{k}$, evaluation of expression $\Ex{e}$ to $\Ex{k}$ is succeeded by execution of command $c\, \Ex{k}$.
\begin{definitionx}[expression-evaluation]
The semantics of the expression evaluation command is defined over the structure of expressions,
as explained below \cite{hayes2021deriving}.
\begin{align}
  \LExpr{v}{lv} & \defs \If \LEx{lv} = \Base{v} \Then \Idle \Else \Magic  \labeldef{lexpr-var} \\
  \LExpr{\Array{lve}{e}}{\Indexed{A}{i}} & \defs \LExpr{lve}{A} \parallel \Expr{e}{i} \labeldef{lexpr-array-indexed} \\
   \LExpr{\Array{lve}{e}}{\Base{v}} & \defs \ \Magic & \mbox{if $\Base{v}$ is a base variable name} \labeldef{lexpr-array-base} \\
  \Expr{\Const{\kappa}}{k} & \defs \If \Ex{k = \kappa} \Then \Idle \Else \Magic  \labeldef{expr-const} \\
  \Expr{\Deref{lve}}{k} & \defs \Nondet_{\LEx{lv}} \spot \LExpr{lve}{lv} \Seq \cgd\Set{\Deref{lv} = k} \Seq \Idle \labeldef{expr-deref} \\
  \Expr{\Unary{\ominus}{e_1}}{k} & \defs \Nondet_{\Ex{k_1}}^{\Ex{k = \opsemantics{\UnaryOp} k_1}} \Expr{e_1}{k_1} \labeldef{expr-unary} \\
  \Expr{\Binary{e_1}{\BinaryOp}{e_2}}{k} & \defs \Nondet_{\Ex{k_1},\Ex{k_2}}^{\Ex{k = k_1 \opsemantics{\BinaryOp} k_2}} (\Expr{e_1}{k_1} \parallel \Expr{e_2}{k_2})  \labeldef{expr-binary} 
\end{align}
\end{definitionx}
For a constant expression $\Const{\kappa}$, the evaluation $\Expr{\Const{\kappa}}{k}$ succeeds if $\Ex{\kappa = k}$ 
but is infeasible otherwise \refdef{expr-const}.
The successful case is represented by the command $\Idle$ and the infeasible case is represented by the command $\Magic$.
When used in an expression of the form \refdef{expr-usage}, 
the only choice that succeeds for $\Expr{\Const{\kappa}}{k}$ is for $\Ex{k} = \Const{\kappa}$.
All other choices for $\Ex{k}$ are infeasible, giving $\Magic$, and so the effect is to execute $c\,\Const{\kappa}$,
allowing first for stuttering steps in the implementation to evaluate the expression and branch to code $c\,\Ex{k}$:
\begin{align*}
  \Nondet_{\Ex{k}} \Expr{\Const{\kappa}}{k} ; c\,\Ex{k} 
  & = \Idle \Seq c\,\Const{\kappa} \nondet \Magic 
     = \Idle \Seq c\,\Const{\kappa}
\end{align*}

Evaluating a variable reference $\Variable{v}$ gives the base variable $\Base{v}$, which is a constant \refdef{lexpr-var}.

Because indexed array references make use of index expressions that typically depend on the state,
we need to define the evaluation of an l-value expression to an l-value (\refdef*{lexpr-array-indexed}--\refdef*{lexpr-array-base}).
For example,
if $\LEx{lv} = \Indexed{v}{i}$,
the l-value expression $\LExpr{\Array{\Base{v}}{e}}{lv}$ simplifies to:
\begin{align*}
  \LExpr{\Array{\Base{v}}{e}}{\Indexed{v}{i}}
= \LExpr{\Base{v}}{v} \parallel \Expr{e}{i}
= \Idle \parallel \Expr{e}{i} 
= \Expr{e}{i}
\end{align*}
and if $\LEx{lv}$ is either a base variable name, or
$\LEx{lv} = \Indexed{lvA}{i}$ where $\LEx{lvA} \neq \Base{v}$, then $\LExpr{\Array{\Base{v}}{e}}{lv} = \Magic$, 
e.g.\ for the case where $\LEx{lv} = \Indexed{lvA}{i}$ and $\LEx{lvA} \neq \LEx{v}$ we have:
\begin{align*}
  \LExpr{\Array{\Base{v}}{e}}{\Indexed{lvA}{i}}
= \LExpr{\Base{v}}{lvA}  \parallel \Expr{e}{i}
= \Magic \parallel \Expr{e}{i} 
= \Magic
\end{align*}
Nested indexed array l-value expressions simplify in an equally straightforward way, e.g.
if $\LEx{lv} = \Indexed{\Indexed{\Base{v}}{i}}{j}$ for any values $\Ex{i}$ and $\Ex{j}$ then
\begin{math}
\LExpr{\Array{\Array{\Base{v}}{e_1}}{e_2}}{lv}
 = \Expr{e_1}{i} \parallel \Expr{e_2}{j}
\end{math}
and
\begin{math}
\LExpr{\Array{\Array{\Base{v}}{e_1}}{e_2}}{lv}
 = \Magic 
\end{math}
otherwise.

Evaluating an expression consisting of a dereference of an l-value expression $\LEx{lve}$ to its r-value $\Ex{k}$, 
$\Expr{\Deref{lve}}{k}$, 
first evaluates the l-value expression to give some l-value $\LEx{lv}$ and 
then checks if the value within state $\sigma$ of $\LEx{lv}$ is $\Ex{k}$,
which may either succeed or become infeasible,
and then allow interference and finite stuttering ($\Idle$) \refdef{expr-deref}.
For example for base variable name $\Base{v}$ we have that 
\begin{align*}
\Expr{ \Deref{\Base{v}} }{k} 
= 
\Nondet_{\LEx{lv}} \spot \LExpr{\Base{v}}{lv} \Seq \cgd\Set{\Deref{lv} = k} \Seq \Idle 
= 
\Idle \Seq \cgd\Set{\Deref{\Base{v}} = k} \Seq \Idle
\end{align*}
holds, and for indexed array reference $\Array{\Base{v}}{e}$ we have that 
\begin{align*}
\Expr{ \Deref{ \Array{\Base{v}}{e} } }{k} 
= 
\Nondet_{\LEx{lv}} \spot \LExpr{\Array{\Base{v}}{e}}{lv} \Seq \cgd\Set{\Deref{lv} = k} \Seq \Idle 
= 
\Nondet_{\Ex{i}} \spot 
       \Expr{e}{i} \Seq \cgd\Set{\Deref{\Indexed{v}{i}} = k} \Seq \Idle
\end{align*}
holds.
The definition of dereferencing an l-value treats access to its r-value as atomic because its value is retrieved in a single state.
For example, it is valid for a 32-bit integer on a 32-bit word machine but not for a 64-bit integer on a 32-bit word machine.

Evaluating a unary expression, $\Expr{\Unary{\ominus}{e_1}}{k}$, 
depends on the evaluation of $\Ex{e_1}$ to some value $\Ex{k_1}$ (i.e.\ $\Expr{e_1}{k_1}$)
such that $\Ex{k = \opsemantics{\ominus} k_1}$,
where $\opsemantics{\ominus}$ is the standard semantics of $\ominus$ on values \refdef{expr-unary}.
For a given value of $\Ex{k}$, 
there may be multiple values of $\Ex{k_1}$ satisfying $\Ex{k = \opsemantics{\ominus} k_1}$, or even no values.
For example, for $\ominus$ as $abs$ (i.e.\ absolute value) 
there are two values for $\Ex{k_1}$ (5 and -5) satisfying $\Ex{5 = abs(k_1)}$ and
no value of $\Ex{k_1}$ satisfying $\Ex{-5 = abs(k_1)}$.
To allow for multiple values of $\Ex{k_1}$ in our definition 
we use a non-deterministic choice over all $\Ex{k_1}$ such that $\Ex{k = \opsemantics{\ominus} k_1}$.
The notation $\Nondet_j^{P\,j} c\,j$, where $P\,j$ is a predicate depending on $j$, is a shorthand for $\Nondet \{ c\,j  \mid j \spot P\,j \}$.
If the unary expression evaluation is in a non-deterministic choice over $\Ex{k}$ and for a particular value of $\Ex{k}$
there is no value of $\Ex{k_1}$ satisfying $\Ex{k = \opsemantics{\UnaryOp} k_1}$ the definition reduces to $\Magic$
and hence is subsumed by alternative values for $\Ex{k}$.

Evaluating a binary expression, $\Expr{\Binary{e_1}{\oplus}{e_2}}{k}$ is defined in similar manner to a unary expression,
with a non-deterministic choice over the possible values $\Ex{k_1}$ and $\Ex{k_2}$ of $\Ex{e_1}$ and $\Ex{e_2}$, respectively,
such that $\Ex{k = k_1 \opsemantics{\oplus} k_2}$, where $\opsemantics{\oplus}$ is the semantics of $\oplus$ on values \refdef{expr-binary}.
Again there may be multiple values of $\Ex{k_1}$ and $\Ex{k_2}$ satisfying $\Ex{k = k_1 \opsemantics{\oplus} k_2}$, or no values.
For example, for $\oplus$ as integer addition, there are multiple values of $\Ex{k_1}$ and $\Ex{k_2}$ satisfying $\Ex{5 = k_1 + k_2}$,
and for $\oplus$ as binary modulo ($\bmod$) there are no values of $\Ex{k_1}$ satisfying $\Ex{6 = k_1 \bmod 5}$.
Because our expressions are free of side effects,
the sub-expressions $\Ex{e_1}$ and $\Ex{e_2}$ can be evaluated in parallel;
that allows arbitrary interleaving of their evaluations.

To handle undefined expressions like $\Binary{\Deref{v}}{\div}{0}$, 
we assume that the semantics of operators on values returns the special value $\UndefinedValue$ for undefined expressions.

From \Definition{expression-evaluation}, 
\refprop{idle-to-expr} and \refprop{idle-to-lexpr} hold because $\Idle = \Idle \Seq \Idle$ and $\Idle = \Idle \parallel \Idle$ 
and $\Nil \refsto \cgd{p}$ for any $p$.
Note that \refprop{idle-to-expr} and \refprop{idle-to-lexpr} hold because expressions are free of side effects.
An $\Idle$ command \refprop{idle-guar-reflexive}, 
an expression evaluation \refprop{expr-guar-reflexive},
and an l-value expression evaluation \refprop{lvalue-guar-reflexive}
all satisfy a reflexive guarantee 
because the only program transitions made by $\Idle$ or an expression evaluation do not change the state.
\begin{align}
  \Idle &\refsto \Expr{e}{k} \labelprop{idle-to-expr} \\
  \Idle &\refsto \LExpr{lve}{lv} \labelprop{idle-to-lexpr} \\
  \guar{g} \together \Idle & = \Idle & \mbox{if $g$ is reflexive} \labelprop{idle-guar-reflexive} \\
  \guar{g} \together \Expr{e}{k} & = \Expr{e}{k} & \mbox{if $g$ is reflexive} \labelprop{expr-guar-reflexive} \\
  \guar{g} \together \LExpr{lve}{lv} & = \LExpr{lve}{lv} & \mbox{if $g$ is reflexive} \labelprop{lvalue-guar-reflexive} 
\end{align}

\subsection{Boolean-valued expressions}\labelsect{expr-boolean}

For boolean-valued expressions, $\Ex{b}$, $\Ex{b_1}$ and $\Ex{b_2}$, and boolean value $k$, 
\begin{align}
  \Expr{\lnot b}{k} & = \Expr{b}{\lnot k} \labelprop{expr-not} \\
  \Expr{\lnot \lnot b}{k} & = \Expr{b}{k} \labelprop{expr-not-not} \\
  \Expr{\lnot(b_1 \land b_2)}{k} & = \Expr{\lnot b_1 \lor \lnot b_2}{k} \labelprop{de-morgan1} \\
  \Expr{\lnot(b_1 \lor b_2)}{k} & = \Expr{\lnot b_1 \land \lnot b_2}{k} \labelprop{de-morgan2}
\end{align}
where to simplify the presentation we have used $\lnot$ for both the syntactic operator
and the semantic operator on boolean values.
Property \refprop{expr-not} follows by expanding the definition of unary not \refdef{expr-unary} as follows:
$\Expr{\lnot b}{k} 
 = \Nondet_{\Ex{k_1}}^{\Ex{k = \lnot k_1}} \Expr{b}{k_1}
 = \Nondet_{\Ex{k_1}}^{\Ex{k1 = \lnot k}} \Expr{b}{\lnot k}
 = \Expr{b}{\lnot k} 
$.
Property \refprop{expr-not-not} follows from \refprop{expr-not}: 
$ \Expr{\lnot \lnot b}{k} 
 = \Expr{\lnot b}{\lnot k} 
 = \Expr{b}{\lnot \lnot k}
 = \Expr{b}{k}
$.
De Morgan's laws, \refprop{de-morgan1} and \refprop{de-morgan2}, follow using \refprop{expr-not} 
and expanding the definition of binary conjunction/disjunction.
Not all laws of boolean algebra apply to boolean expressions.
For example, we do not have a law of the form,
$\Expr{b_1 \lor (b_2 \land b_3)}{k} = \Expr{(b_1 \lor b_2) \land (b_1 \lor b_3)}{k}$,
because it introduces multiple occurrences of $\Ex{b_1}$ on the right.%
\footnote{Note that we do have $\Expr{(b_1 \lor b_2) \land (b_1 \lor b_3)}{k} \refsto \Expr{b_1 \lor (b_2 \land b_3)}{k}$
because the two occurrence of $\Ex{b_1}$ on the left may evaluate to the same value.}

\subsection{Infeasible expression evaluations}\labelsect{expr-infeasible}

An expression evaluation  $\Expr{e}{k}$ is infeasible if $\Expr{e}{k} = \Expr{e}{k} \Seq \Magic$.
For example, a constant expression evaluation $\Expr{\kappa}{k}$ is infeasible if $\Ex{\kappa \neq k}$.
A unary expression evaluation, $\Expr{\UnaryOp e_1}{k}$, is infeasible 
if there is no value $\Ex{k_1}$ satisfying $\Ex{k = \opsemantics{\UnaryOp} k_1}$,
and $\Expr{e_1 \BinaryOp e_2}{k}$ is infeasible
if there are no values $\Ex{k_1}$ and $\Ex{k_2}$ satisfying $\Ex{k = k_1 \opsemantics{\BinaryOp} k_2}$.
Note that an infeasible expression evaluation is typically in the context of a non-deterministic choice over $\Ex{k}$,
and $\Expr{e}{k}$ not being feasible means that choice of $\Ex{k}$ can effectively be ignored 
because it will be subsumed by a feasible choice.
If $\Expr{e}{k}$ is infeasible, 
then it achieves \emph{any} post-condition $p_1$ in a Hoare triple:
\begin{align}
  \Triple{p}{\rely{r} \together \Expr{e}{k}}{p_1}  \labelprop{non-realisable-eval-post}
\end{align}
because $\Triple{p}{{\rely{r}} \together \Expr{e}{k}\Seq\Magic}{p_1}$ holds for any $p_1$
because $\Magic$ is a left annihilator of sequential composition.

\section{Reasoning compositionally about expression evaluation}\labelsect{complex-expressions}

\labelsect{motivation}
Consider the following conditional command,
\begin{align}
  \Pre{p} \Seq \If b \Then \Pre{p_t} \Seq c \Else \Pre{p_f} \Seq d \Fi . \labelprop{if-pt-pf}
\end{align}
For a sequential program \cite{Hoare69a}, 
assuming the initial state satisfies the precondition $p$,
the assertions $p_t$ and $p_f$ in the conditional 
hold if $p_t$ is taken to be $p \inter b$ and $p_f$ is taken to be $p \inter \lnot b$.
However, under interference from concurrent threads, 
the choices for $p_t$ and $p_f$ need to be adapted to 
account for possible interference from concurrent threads during the evaluation of the guard $b$ because:
\begin{enumerate}
\item\labelprop{interference-pre}
the precondition $p$ may be invalidated by interference,
\item\labelprop{interference-true}
if $b$ evaluates to $\True$, it does not follow that $b$ holds at the start of the $\Then$ branch,
and 
\item\labelprop{interference-false}
if $b$ evaluates to $\False$, it does not follow that $\lnot b$ holds at the start of the $\Else$ branch.
\end{enumerate}
In the context of a rely condition $r$,
the precondition $p$ will not be invalidated if $p$ is stable under $r$.
\begin{definitionx}[stable]
A set of states, $p$, is \emph{stable} under a relation $r$ if, whenever $\sigma \in p$ and $(\sigma,\sigma') \in r$, then $\sigma' \in p$.
\end{definitionx}

For reasoning about guard evaluation under interference satisfying the rely condition $r$, we make use of a Hoare triple (\Definition*{Hoare-triple})
of the form, $\Establish{p}{r}{c}{p_1}$,
which can be read as $c$ establishes $p_1$ from initial states satisfying $p$ under interference satisfying $r$.
The command $\Idle$ \refdef{idle} is the most general command that only performs a finite number of program transitions
all of which do not change the state and also allows any environment transitions, and so satisfies \refprop{idle-stable}
because if $p$ is stable under $r$,
then $p$ will hold after interference for which every transition satisfies $r$.
Because expressions refine $\Idle$ by \refprop{idle-to-expr} they satisfy \refprop{expr-idle-stable}.
\begin{align}
\Triple{p}{\rely{r} \together \Idle}{p} && \mbox{if $p$ is stable under $r$} \labelprop{idle-stable} \\
\Triple{p}{\rely{r} \together \Expr{e}{k}}{p} && \mbox{if $p$ is stable under $r$} \labelprop{expr-idle-stable}
\end{align}

\begin{RelatedWork}
Jones \cite{Jones81d} made use of a similar property that invariants are preserved by $r$.
Our definition of stability corresponds to that of Xu et al.\ \cite{XuRoeverHe97}.
Wickerson et al.\ \cite{DBLP:conf/esop/WickersonDP10}, as well as defining stability as in \Definition*{stable},
define
$\lfloor p \rfloor_r$ for the weakest assertion that is stronger than $p$ and stable under $r$, and 
$\lceil p \rceil_r$ for the strongest assertion that is weaker than $p$ and stable under $r$,
which are analogous to Dijkstra's weakest/strongest pre/post-conditions \cite{Dijkstra75,Dijkstra76}.
Such ideas could be applied within our theory but we have not found it necessary.
\end{RelatedWork}

If we consider the conditional command in \refprop{if-pt-pf} executing in a context 
in which its environment satisfies a rely condition $r$,
the assertions $p_t$ and $p_f$ in \refprop{if-pt-pf} will be valid if the following Hoare triples hold
because these triples correspond to evaluating the guard to $\True$/$\False$ under interference satisfying $r$.
\begin{align}
  \EstablishExpr{p}{r}{b}{\True}{p_t} \labelprop{establish-pt} \\
  \EstablishExpr{p}{r}{b}{\False}{p_f} \labelprop{establish-pf}
\end{align}
For example, if $p$ is stable under $r$, 
then as a special case properties \refprop{establish-pt} and \refprop{establish-pf} are guaranteed to hold if $p_t = p$ and $p_f = p$. 
To show that the assertions $p_t$ and $p_f$ hold in the while loop,
\begin{align}
  \Pre{p} \Seq \While b \Do \Pre{p_t} \Seq c \Od \Seq \Pre{p_f}  \labelprop{while-inference}
\end{align}
the same triples \refprop{establish-pt} and \refprop{establish-pf} can
be used, with the assumption that $p$ is both stable under $r$
and an additional requirement that $p$ is an invariant of the loop, that is, if $p$
holds initially, $p$ holds after each iteration.

Our motivation above focused on boolean expressions
but below we develop more general laws to show the validity of a Hoare triple of the form, 
$\EstablishExpr{p}{r}{e}{k}{P\,k}$, for an expression of any type,
where we use $P\,k$ for the postcondition because the postcondition usually depends on $k$.
Such laws can be used for both the expressions in assignment commands and 
the boolean guards in conditionals and while loops
(see \refsect{commands}).
Because expression evaluation satisfies any reflexive guarantee \refprop{expr-guar-reflexive}, 
we do not need consider guarantees explicitly.

To reason about arbitrary sets of states
and arbitrary expressions, possibly with multiple references to shared variables,
we make use of a compositional approach in which we establish a postcondition for a complex expression 
by first establishing postconditions for each of its component subexpressions.
The approach does not require any syntactic side conditions about the form of an expression and its use of shared variables.
The following subsections develop laws for inductively reasoning about each of the forms of expressions.

\subsection{Hoare triples for constant expressions}\labelsect{Hoare-constant}

The base case for expressions is a constant expression $\kappa$.
\begin{lemmax}[post-constant]
For values $\Const{\kappa}$ and $\Const{k}$, a relation $r$, a set of states $p$ that is stable under $r$ the following triple holds,
$\EstablishExpr{p}{r}{\Const{\kappa}}{k}{p \inter \Set{\Const{\kappa} = \Const{k}}}$.
\end{lemmax}

\begin{proof}
If $\Const{\kappa} \neq \Const{k}$, the expression evaluation is infeasible and the triple holds by \refprop{non-realisable-eval-post}.
If $\Const{\kappa} = \Const{k}$, the triple holds by \refprop{expr-idle-stable}
because $p$ is stable under $r$.
\end{proof}

\subsection{Hoare triples for l-value expressions}\labelsect{Hoare-lvalue}

The base case is for an l-value expression that is a variable $\Variable{v}$,
which is independent of the state and evaluates to $\LEx{lv}$ if and only of $\Variable{v} = \LEx{lv}$.
\begin{lemmax}[post-variable]
For a variable $\Variable{v}$, an l-value $\LEx{lv}$, a relation $r$, and a set of states $p$ that is stable under $r$, 
$\EstablishLExpr{p}{r}{v}{lv}{p \inter \Set{\Variable{v} = \LEx{lv}}}$.
\end{lemmax}

\begin{proof}
If $\Variable{v} \neq \LEx{lv}$, the expression evaluation is infeasible and the triple holds by \refprop{non-realisable-eval-post}.
If $\Variable{v} = \LEx{lv}$, the triple holds by \refprop{expr-idle-stable}
because $p$ is stable under $r$.
\end{proof}

To evaluate an array reference l-value expression $\Array{lve}{e}$,
one needs to evaluate both the l-value expression $\LEx{lve}$ to some l-value $\LEx{A}$ and 
the index expression $\Ex{e}$ to some value $\Ex{i}$.
The evaluation $\LExpr{lve}{A}$ is assumed to establish $P_1\,\LEx{A}$ and
the evaluation $\Expr{e}{i}$ is assumed to establish $P_2\,\Ex{i}$.
Together they establish $P\,\LEx{lv}$, where $\LEx{lv} = \Indexed{A}{i}$.
\begin{lemmax}[post-array-ref]
For 
a reflexive relation $r$, 
a set of states $p$, 
an l-value expression $\LEx{lve}$,
an expression $\Ex{e}$,
a function $P_1$ from l-values to sets of states,
and
a function $P_2$ from values to sets of states,
if,
\begin{align}
  \forall \LEx{A} \spot  \Triple{p}{&\rely{r} \together \LExpr{lve}{A}}{P_1\,\LEx{A}} \labelprop{array-lve} \\
  \forall \Ex{i} \spot \Triple{p}{&\rely{r} \together \Expr{e}{i}}{P_2\,\Ex{i}} \labelprop{array-e} \\
  \forall \LEx{A}\,\Ex{i} \spot \LEx{lv} = \Indexed{A}{i} &\implies P_1\,\LEx{A} \inter P_2\,\Ex{i} \subseteq P\,\LEx{lv} \labelprop{array-P}
\end{align}
then,
$\Triple{p}{\rely{r} \together \LExpr{\Array{lve}{e}}{lv}}{P\,\LEx{lv}}$.
\end{lemmax}

\begin{proof}
For the case where $ \LEx{lv} = \LEx{\Indexed{A}{i}}$ we have from
\refdef{lexpr-array-indexed} that $\LExpr{\Array{lve}{e}}{lv} =
\LExpr{lve}{A} \parallel \Expr{e}{i}$, and
$$\Triple{p}{\rely{r} \together (\LExpr{lve}{A} \parallel \Expr{e}{i})}{P\,\LEx{lv}}$$
holds by Hoare inference rules \refprop{hoare-parallel} and \refprop{hoare-consequence} using
using assumptions \refprop{array-lve}--\refprop{array-P}.
For the case where $\LEx{lv} = \Base{v}$ is a base variable name, we have
from \refdef{lexpr-array-base} that $\LExpr{\Array{lve}{e}}{lv} = \Magic$,
and hence the triple holds by \refprop{non-realisable-eval-post}.
\end{proof}

\subsection{Hoare triples for dereferencing an l-value expression}\labelsect{Hoare-deref}

A dereference of an l-value expression $\Deref{lve}$ first evaluates $\LEx{lve}$ to get an l-value $\LEx{lv}$
and then atomically dereferences $\LEx{lv}$ to get the value of the expression $\Ex{k}$. 
\begin{lemmax}[post-deref-lvalue]
For 
a relation $r$,
a set of states $p$ that is stable under $r$,
an l-value expression $\LEx{lve}$,
and
a function $P_1$ from l-values to set of states,
and
a function $P$ from values to sets of states that are stable under $r$,
if for all l-values $\LEx{lv}$ both,
\begin{align}
  & \Triple{p}{\rely{r} \together \LExpr{lve}{lv}}{P_1\,\LEx{lv}} \labelprop{deref-establish-P1} \\
  & P_1\,\LEx{lv} \inter \Set{\Deref{lv} = \Ex{k}} \subseteq P\,\Ex{k} \labelprop{deref-establish-P}
\end{align}
then
$\Triple{p}{\rely{r} \together \Expr{\Deref{lve}}{k}}{P\,\Ex{k}}$. 
\end{lemmax}

\begin{proof}
From the definition of dereferencing an l-value expression \refdef{expr-deref}, 
it is sufficient to show:
\begin{align}
\Triple{p}{
  \rely{r} \together \Nondet_{\LEx{lv}} \spot \LExpr{lve}{lv} \Seq \cgd\Set{\Deref{lv} = k} \Seq \Idle
}
{P\,\Ex{k}}
\labelprop{conclusion-post-deref-lvalue}
\end{align}
From assumption \refprop{deref-establish-P} and property \refprop{assert-plus-test}
we have that, for all $lv$:
\begin{align}
& \Triple{P_1\,\LEx{lv}}{\rely{r} \together \cgd\Set{\Deref{lv} = k} }{P\,\Ex{k}}
  \labelprop{lookup-establish-P}
\end{align}
and from the assumption that function $P\,k$ is stable under $r$, and property \refprop{idle-stable} we have that
\begin{align}
& \Triple{P\,\Ex{k}}{\rely{r} \together \Idle}{P\,\Ex{k}} \labelprop{idle-preserve-P}
\end{align}
Applying the Sequence Hoare inference rule \refprop{hoare-sequential} twice 
using \refprop{deref-establish-P1}, \refprop{lookup-establish-P} and \refprop{idle-preserve-P}, 
we can deduce,
\begin{align}
  \Triple{p}{\rely{r} \together \LExpr{lve}{lv} \Seq \cgd\Set{\Deref{lv} = k} \Seq \Idle}{P\,k}
\end{align}
and finally \refprop{conclusion-post-deref-lvalue} holds by the Non-deterministic Choice Hoare inference rule \refprop{hoare-Nondet}.
\end{proof}

The following corollary applies \reflem{post-deref-lvalue} to the special case 
where the l-value is a variable name (using \reflem{post-variable}).
\begin{corx}[post-deref-variable]
For a relation $r$,
set of states $p$ that is stable under $r$,
variable name $\Variable{v}$, 
and
function $P$ from values to sets of states that are stable under $r$,
if,
\begin{align}
  p \inter \Set{\Deref{v} = \Ex{k}} \subseteq P\,\Ex{k} 
\end{align}
then $\Triple{p}{\rely{r} \together \Expr{\Deref{v}}{k}}{P\,\Ex{k}}$.
\end{corx}

\begin{example}\label{ex-vars}
For atomic access variables $\Variable{v}$ and $\Variable{u}$ and constants $\Ex{k_1}$ and $\Ex{k_2}$,
because $\EqEval{\Deref{v}}{k_1} \subseteq \LEEval{\Deref{v}}{k_1}$ and $ \LEEval{\Deref{v}}{k_1}$ is 
stable under a rely condition that does not increase $\Variable{v}$,
one can deduce \refprop{vars-v} by \refcor{post-deref-variable}.
Similarly, under a rely condition that does not decrease $\Variable{u}$, one can deduce \refprop{vars-u}. 
\begin{align}
  & \EstablishExpr{\Set{\True}}{\Rel{\Deref{v}' \leq \Deref{v}}}{\Deref{v}}{k_1}{\Set{\Deref{v} \leq k_1}}  \labelprop{vars-v} \\
  & \EstablishExpr{\Set{\True}}{\Rel{\Deref{u} \leq \Deref{u}'}}{\Deref{u}}{k_2}{\Set{k_2 \leq \Deref{u}}}  \labelprop{vars-u} 
\end{align}
\end{example}

\begin{example}\label{negate-var}
If interference from the environment can only negate a variable $\Variable{v}$ or leave $\Variable{v}$ unchanged,
that is $r \subseteq \Rel{\Deref{v}' = -\Deref{v} \lor \Deref{v}' = \Deref{v}}$,
\begin{align*}
  \EstablishExpr{\Set{\True}}{r}{\Deref{v}}{k}{\Set{\Deref{v} = \Ex{k} \lor -\Deref{v} = \Ex{k}}} .
\end{align*}
\end{example}

\subsection{Hoare triples for unary expressions}\labelsect{Hoare-unary}

To show that the unary expression $\Expr{\Unary{\UnaryOp}{e_1}}{k}$ establishes postcondition $P\,\Ex{k}$ 
from initial states satisfying $p$ under interference satisfying $r$,
a similar property that $\Expr{e_1}{k_1}$ establishes $P_1\,\Ex{k_1}$ is required for the subexpression $\Ex{e_1}$. 
\begin{lemmax}[post-unary]
Given an expression $\Ex{e_1}$,
a value $\Const{k}$,
a relation $r$,
a set of states $p$,
and functions $P_1$ and $P$ from values to sets of states,
if for all $\Const{k_1}$ such that $\Ex{k = \opsemantics{\UnaryOp} k_1}$ both,
\begin{align}
  & \EstablishExpr{p}{r}{e_1}{k_1}{P_1\,\Ex{k_1}}  \labelprop{post-unary1} \\
  & P_1\,\Ex{k_1} \subseteq P\,\Ex{k} \labelprop{post-unary-promote}
\end{align}
then, 
$\EstablishExpr{p}{r}{\Unary{\UnaryOp}{e_1}}{k}{P\,\Ex{k}}$.
\end{lemmax}

\begin{proof}
If there does not exist a value of $\Ex{k_1}$ such that $\Ex{k = \opsemantics{\ominus} k_1}$, 
the expression evaluation $\Expr{\Unary{\UnaryOp}{e_1}}{k}$ is infeasible, the result trivially holds by \refprop{non-realisable-eval-post},
otherwise, 
using assumptions \refprop{post-unary1} and \refprop{post-unary-promote}, 
we can infer using \refprop{hoare-consequence} that for all $\Ex{k_1}$ such that $\Ex{k = \opsemantics{\UnaryOp} k_1}$,
\begin{align}
  \EstablishExpr{p}{r}{e_1}{k_1}{P\,\Ex{k}}
\end{align}
holds, and using \refprop{hoare-Nondet} we can infer,
\begin{align}
  \EstablishExpr{p}{r}{\Nondet_{k_1}^{k = \opsemantics{\ominus} k_1} \Expr{e_1}{k_1}}{k}{P\,k}
\end{align}
and the lemma follows from the definition of $\Expr{\ominus e_1}{k}$ \refdef{expr-unary}.
\end{proof}

\begin{example}
If interference on a thread can only negate a variable $\Variable{v}$ or leave it unchanged, 
that is, $r \subseteq \Rel{\Deref{v}' = -\Deref{v} \lor \Deref{v}' = \Deref{v}}$, from Example~\ref{negate-var}, we have,
\[
  \EstablishExpr{\Set{\True}}{r}{\Deref{v}}{k_1}{\Set{\Deref{v} = k_1 \lor -\Deref{v} = k_1}}
\]
and if $\Ex{k = abs(k_1)}$ then $\Deref{v} = \Ex{k_1} \lor -\Deref{v} = \Ex{k_1} \implies \Ex{k = abs(\Deref{v})}$,
and hence by \reflem{post-unary}
\[
  \EstablishExpr{\True}{r}{abs(\Deref{v})}{k}{\Set{k = abs(\Deref{v})}} .
\]
Note that, although $\Variable{v}$ may be modified by interference satisfying $r$, 
the value of $\Ex{abs(\Deref{v})}$ remains unchanged by interference satisfying $r$.
\end{example}

\subsection{Hoare triples for binary expressions}\labelsect{Hoare-binary}

To show that $\Expr{\Binary{e_1}{\BinaryOp}{e_2}}{k}$ establishes postcondition $P\,\Ex{k}$ from initial states satisfying $p$ under interference satisfying $r$,
similar properties are required for the sub-expressions $\Ex{e_1}$ and $\Ex{e_2}$. 

\begin{lemmax}[post-binary]
Given 
expressions $\Ex{e_1}$ and $\Ex{e_2}$,
a value $\Ex{k}$,
a reflexive relation $r$,
a set of states $p$,
and
functions $P_1$, $P_2$ and $P$ from values to sets of states,
if for all $\Ex{k_1}$ and $\Ex{k_2}$ such that $\Ex{k = k_1 \opsemantics{\BinaryOp} k_2}$ the following three properties hold,
\begin{align}
  & \EstablishExpr{p}{r}{e_1}{k_1}{P_1\,\Ex{k_1}}  \labelprop{post-binary1} \\
  & \EstablishExpr{p}{r}{e_2}{k_2}{P_2\,\Ex{k_2}}  \labelprop{post-binary2} \\
  & P_1\,\Ex{k_1} \inter P_2\,\Ex{k_2} \subseteq P\,\Ex{k} \labelprop{post-binary-promote}
\end{align}
then,
$\EstablishExpr{p}{r}{\Binary{e_1}{\BinaryOp}{e_2}}{k}{P\,\Ex{k}}$.
\end{lemmax}

\begin{proof}
If there do not exist values of $\Ex{k_1}$ and $\Ex{k_2}$ such that $\Ex{k = k_1 \opsemantics{\oplus} k_2}$, 
the expression evaluation $\Expr{\Binary{e_1}{\BinaryOp}{e_2}}{k}$ is infeasible, the result trivially holds by \refprop{non-realisable-eval-post},
otherwise, because $r$ is reflexive by \refprop{expr-guar-reflexive} both 
$\guar{r} \together \Expr{e_1}{k_1} = \Expr{e_1}{k_1}$
and
$\guar{r} \together \Expr{e_2}{k_2} = \Expr{e_2}{k_2}$,
and hence using the Parallel Hoare inference rule \refprop{hoare-parallel} with assumptions \refprop{post-binary1} and \refprop{post-binary2} 
one can deduce that for all $k$ such that $k = k_1 \oplus k_2$, 
\begin{align*}&
  \Triple{p}{\rely{r} \together (\Expr{e_1}{k_1} \parallel \Expr{e_2}{k_2})}{P_1\,k_1 \inter P_2\,k_2}
 \Entails*[by the Consequence Hoare inference rule \refprop{hoare-consequence} using \refprop{post-binary-promote}]
  \Triple{p}{\rely{r} \together (\Expr{e_1}{k_1} \parallel \Expr{e_2}{k_2})}{P\,k}
\end{align*}
and hence, using the Non-deterministic Choice Hoare inference rule \refprop{hoare-Nondet},
\begin{align*}&
  \Triple{p}{\rely{r} \together \Nondet_{k_1,k_2}^{k = k_1 \opsemantics{\oplus} k_2} (\Expr{e_1}{k_1} \parallel \Expr{e_2}{k_2})}{P\,k}
\end{align*}
and hence the lemma follows by the definition of a binary expression evaluation \refdef{expr-binary}.
\end{proof}

\begin{example}\label{ex-v-le-u}
For atomic access variables $\Variable{v}$ and $\Variable{u}$,
if rely condition $r$ ensures $\Variable{v}$ does not increase and $\Variable{u}$ does not decrease,
(i.e.\ $r  \subseteq \Rel{\Deref{v}' \leq \Deref{v} \land \Deref{u} \leq \Deref{u}'}$)
because for all $\Ex{k_1}$ and $\Ex{k_2}$ such that $\Ex{k_1 \leq k_2}$ we have, 
$\LEEval{\Deref{v}}{k_1} \inter \LEEval{k_2}{\Deref{u}} \subseteq \LEEval{\Deref{v}}{\Deref{u}}$,
we can deduce \refprop{leq-true} using \reflem{post-binary} using the results from Example~\ref{ex-vars}.
For the case where the expression $\Ex{\Deref{v} \leq \Deref{u}}$ evaluates to $\False$, 
one can only conclude that $\Set{\True}$ holds \refprop{leq-false},
which gives no additional information.
\begin{align}
  & \EstablishExpr{\Set{\True}}{\Rel{\Deref{v}' \leq \Deref{v} \land \Deref{u} \leq \Deref{u}'}}{\Deref{v} \leq \Deref{u}}{\True}{\Set{\Deref{v} \leq \Deref{u}}}  \labelprop{leq-true} \\
  & \EstablishExpr{\Set{\True}}{\Rel{\Deref{v}' \leq \Deref{v} \land \Deref{u} \leq \Deref{u}'}}{\Deref{v} \leq \Deref{u}}{\False}{\Set{\True}}  \labelprop{leq-false} 
\end{align}
\end{example}

\begin{example}
If interference can only modify an integer variable $\Variable{v}$ by adding a multiple of two,
that is, $r \subseteq \Set{even(\Deref{v}' - \Deref{v})}$,
then 
\[
  \EstablishExpr{\Set{\True}}{r}{\Deref{v} \bmod 2}{k}{\Set{k = \Deref{v} \bmod 2}}
\]
holds by \reflem{post-binary} because $\Ex{even(\Deref{v}' - \Deref{v}) \implies (\Deref{v} \bmod 2 = \Deref{v}' \bmod 2)}$
and hence $\Ex{(k = \Deref{v} \bmod 2)}$ is stable under $r$ and 
is  implied by $\Ex{k = k_1 \bmod 2 \land k_1 = \Deref{v}}$.
Note that, although $\Variable{v}$ may be modified by interference satisfying $r$,
the value of $\Ex{\Deref{v} \bmod 2}$ remains unchanged by interference satisfying $r$.
\end{example}

\subsection{Expressions with no variables subject to interference}\labelsect{Hoare-no-interference}

Local variables and shared variables ``owned'' by a thread are not subject to interference 
and hence their values are invariant under the the thread's rely condition.
We define invariance under a rely condition for an arbitrary expression.
An expression $\Ex{e}$ is \emph{invariant} under a relation $r$ if and only if 
all dereferences of l-values within $\Ex{e}$ are invariant under $r$.
\begin{definitionx}[invar-expr/invar-lval]
An expression is invariant under a rely condition $r$, if each of its sub-expressions (including l-value sub-expressions) is invariant.
\begin{align}
  \textsf{invar-expr}\,\Const{\kappa}\,r & = \True \labelprop{invar-const} \\
  \textsf{invar-expr}\,(\Unary{\UnaryOp}{e_1})\,r & = \textsf{invar-expr}\,\Ex{e_1}\,r \labelprop{invar-unary} \\
  \textsf{invar-expr}\,(\Binary{e_1}{\BinaryOp}{e_2})\,r & = \textsf{invar-expr}\,\Ex{e_1}\,r \land \textsf{invar-expr}\,\Ex{e_2}\,r \labelprop{invar-binary} \\
  \textsf{invar-expr}\,(\Deref{lve})\,r & = \textsf{invar-lval}\,\LEx{lve} \land 
  					(\forall (\sigma,\sigma') \in r \spot get\,(\LEval{lve}{\sigma})\,\sigma = get\,(\LEval{lve}{\sigma'})\,\sigma'  \labelprop{invar-deref} \\
  \textsf{invar-lval}\,\Variable{v}\,r & = \True \labelprop{invar-var} \\
  \textsf{invar-lval}\,\Array{lve}{e_1}\,r & = \textsf{invar-lval}\,\LEx{lve}\,r \land \textsf{invar-expr}\,\Ex{e_1}\,r \labelprop{invar-array}
\end{align}

\end{definitionx}

A special case of \refcor{post-deref-variable} is if the variable access $\Deref{\Variable{v}}$ is invariant (unmodified) under $r$,
in which case $\Set{\Deref{v} = k}$ is stable under $r$.
\begin{corx}[post-variable-invariant]
Given a variable $\Variable{v}$, 
a relation $r$, 
and 
a set of states $p$ that is stable under $r$, 
if $\Deref{\LEx{v}}$ is invariant under $r$ then, $\EstablishExpr{p}{r}{\Deref{v}}{k}{p \inter \Set{\Deref{v} = k}}$.
\end{corx}

More generally if an expression $\Ex{e}$ is invariant under $r$ then evaluating $\Expr{e}{k}$ in the context of a rely condition $r$ will establish $\Set{e = k}$.
\begin{lemmax}[post-expr-invariant]
Given an expression $e$ (or l-value expression $lve$), 
a relation $r$, 
and 
a set of states $p$ that is stable under $r$,
if $e$ is invariant under $r$ then,
$\EstablishExpr{p}{r}{e}{k}{p \inter \Set{e = k}}$.
\end{lemmax}

\begin{proof}
The proof is by induction over the structure of an expression. 
The base cases for $\Ex{e}$ either a constant $\kappa$ or a variable $\Variable{v}$ follow from \reflem{post-constant} and
\refcor{post-variable-invariant}, respectively.

For the unary expression $\Unary{\ominus}{e_1}$, 
$\Ex{e_1}$ is invariant under $r$ by \refprop{invar-unary} and hence we have the inductive hypothesis,
$\EstablishExpr{p}{r}{e_1}{k_1}{p \inter \Set{e_1 = k_1}}$.
Now $\Ex{k = \ominus k_1 \land \Ex{e_1} = k_1}$ implies $\Ex{\ominus e_1} =k$,
and hence using \reflem{post-unary} we can deduce,
$\EstablishExpr{p}{r}{\Unary{\ominus}{e_1}}{k}{p \inter \Set{\Unary{\ominus}{e_1} = k}}$ as required.

For the binary expression $\Binary{e_1}{\oplus}{e_2}$, 
both $\Ex{e_1}$ and $\Ex{e_2}$ are invariant under $r$ by \refprop{invar-binary} and hence we have the inductive hypotheses,
$\EstablishExpr{p}{r}{e_1}{k_1}{p \inter \Set{e_1 = k_1}}$ and
$\EstablishExpr{p}{r}{e_2}{k_2}{p \inter \Set{e_2 = k_2}}$.
Now $\Ex{k = k_1 \oplus k_2 \land \Ex{e_1} = k_1 \land \Ex{e_2} = k_2}$ implies $\Binary{e_1}{\oplus}{e_2} = k$,
and hence using \reflem{post-binary} we can deduce,
$\EstablishExpr{p}{r}{\Binary{e_1}{\oplus}{e_2}}{k}{p \inter \Set{\Binary{e_1}{\oplus}{e_2} = k}}$ as required.

For the dereference expression $\Deref{lve}$,
$\LEx{lve}$ is invariant under $r$ by \refprop{invar-deref} and hence we can assume the inductive hypothesis,
$\EstablishLExpr{p}{r}{lve}{lv}{p \inter \Set{\LEx{lve} = \LEx{lv}}}$.
Now we have that $\Ex{k} = \Deref{lv} \land \LEx{lve} = \LEx{lv}$ implies $\Deref{lve} = \Ex{k}$ 
and hence using \reflem{post-deref-lvalue} we can deduce the Hoare triple,
$\EstablishExpr{p}{r}{\Deref{lve}}{k}{p \inter \Set{\Deref{lve} = k}}$ 
because $\Set{\Deref{lve} = k}$ is stable under $r$ by \refprop{invar-deref}.

For the l-value expression $\Array{lve}{e_1}$, 
both $\LEx{lve}$ and $\Ex{e_1}$ are invariant under $r$ by \refprop{invar-array} and hence we have the inductive hypotheses,
$\EstablishLExpr{p}{r}{lve}{A}{p \inter \Set{\LEx{lve} = \LEx{A}}}$ and
$\EstablishExpr{p}{r}{e_1}{i}{p \inter \Set{e_1 = i}}$.
Now $\LEx{lv} = \LEx{\Indexed{A}{i}} \land \LEx{lve} = \LEx{A} \land \Ex{e_1 = i}$ implies $\LEx{lv} = \LEx{\Indexed{lve}{e_1}}$,
and hence using \reflem{post-array-ref} we can deduce,
$\EstablishLExpr{p}{r}{\Array{lve}{e_1}}{lv}{p \inter \Set{\Array{lve}{e_1} = \LEx{lv}}}$ as required.
\end{proof}

\begin{RelatedWork}
\reflem{post-expr-invariant} 
formalises a rule used informally in earlier work
\cite{Jones81d,Jones83a,Jones83b,Stolen90,stolen1991method,XuRoeverHe97,Dingel02,PrensaNieto03,Wickerson10-TR,DBLP:conf/esop/WickersonDP10,SchellhornTEPR14,Sanan21}.
Note that those approaches all use an operational semantics in which expression evaluation is treated as atomic
and hence their theories cannot be used to reason about non-atomic expression evaluation.
Instead they make use of the restriction that each expression can have only a single critical reference. 
The validity of this restriction cannot be proven within their theories because of the atomicity assumptions in their operational semantics.
If an expression does have a single critical reference, 
all sub-expressions not containing the critical reference can be reasoned about using \reflem{post-expr-invariant} 
and the remaining sub-expressions can use the more general rules above.
\end{RelatedWork}

The following lemma handles the special case of an array reference l-value 
for a one-dimensional array indexed with an expression that is invariant under $r$.
\begin{lemmax}[post-array-invar-index]
For a reflexive relation $r$,
set of states $p$ that is stable under $r$,
a single-dimensional array variable $\Variable{v}$,
and
index expression $\Ex{e}$ that is invariant under $r$,
if 
\begin{align}
  \forall \Ex{i} \spot \LEx{lv} = \Indexed{v}{i} \implies p \inter \Set{e = i} \subseteq P\,\LEx{lv} \labelprop{single-invar}
\end{align}
then 
$\EstablishLExpr{p}{r}{\Array{\Variable{v}}{e}}{lv}{P\,\LEx{lv}}$.
\end{lemmax}

\begin{proof}
Using \reflem{post-variable} and \reflem{post-expr-invariant} we have both,
\begin{align}
  \forall \LEx{A} \spot \Triple{p}{&\rely{r} \together \LExpr{\Variable{v}}{A}}{p \inter \Set{\Base{v} = \LEx{A}}} \labelprop{single-A} \\
  \forall \Ex{i} \spot \Triple{p}{&\rely{r} \together \Expr{e}{i}}{p \inter \Set{e = i}} \labelprop{single-e}
\end{align}
and using \refprop{single-invar} gives,
\begin{align}
  \forall \LEx{A}\,\Ex{i} \spot \LEx{lv} = \LEx{\Indexed{A}{i}} \implies p \inter \Set{\Base{v} = \LEx{A} \land e = i} \subseteq P\,\LEx{lv} 
\end{align}
and the lemma follows using \reflem{post-array-ref}.
\end{proof}

While the above lemma handles evaluating an l-expression for a single-dimensional array with an index expression that is invariant under $r$,
the following lemma handles dereferencing such an l-value expression.
Note that the value of the array element may be subject to interference.
\begin{lemmax}[post-array-invar-index-deref]
For a reflexive relation $r$,
a set of states $p$ that is stable under $r$,
a single-dimensional array variable $\Variable{v}$,
index expression $\Ex{e}$ that is invariant under $r$,
and
function $P$ from values to sets of states that are stable under $r$,
if,
\begin{align}
  p \inter \Set{\Deref{(\Array{\Variable{v}}{e})} = k} \subseteq P\,\Ex{k} \labelprop{single-deref-P}
\end{align}
then
$\EstablishExpr{p}{r}{\Deref{(\Array{\Variable{v}}{e})}}{k}{P\,\Ex{k}}$.
\end{lemmax}

\begin{proof}
Because we have a single-dimensional array and an index expression that is invariant under $r$,
the array reference l-value is invariant under $r$ and hence,
\begin{align}
  \forall \LEx{lv} & \spot stable\,(p \inter \Set{\Array{\Variable{v}}{e} = \LEx{lv}})\,r \labelprop{single-deref-stable-P1} \\
  \forall \Ex{i}\,\LEx{lv} & \spot \LEx{lv} = \LEx{\Indexed{v}{i}} \implies p \inter \Set{e = i} \subseteq p \inter \Set{\Array{\Variable{v}}{e} = \LEx{lv}} \labelprop{single-deref-Plv} 
\end{align}
and hence using \reflem{post-array-invar-index} we have,
\begin{align}
  \forall \LEx{lv} \spot \Triple{p}{\rely{r} \together \LExpr{\Array{\Variable{v}}{e}}{lv}}{p \inter \Set{\Array{\Variable{v}}{e} = \LEx{lv}}} \labelprop{single-deref-P1}
\end{align}
and also by assumption \refprop{single-deref-P}
\begin{align}
  \forall \LEx{lv} \spot p \inter \Set{\Array{\Variable{v}}{e} = \LEx{lv} \land \Deref{lv} = k} \subseteq P\,\Ex{k}
\end{align}
from which the lemma follows using \reflem{post-deref-lvalue}.
\end{proof}

\subsection{Laws for boolean operations}\labelsect{Hoare-boolean}

Boolean conjunction and disjunction are important special cases of binary operators.
In this section we develop laws for conjunction and disjunction based on \reflem{post-binary}.
\reflem{post-conjoin} establishes a postcondition, $p_1 \inter p_2$, for the boolean expression $b_1 \land b_2$,
based on the assumptions that $b_1$ establishes the postcondition $p_1$ and $b_2$ establishes the postcondition $p_2$.
\begin{lemmax}[post-conjoin]
If $\Ex{b_1}$ and $\Ex{b_2}$ are boolean expressions,
$r$ is a reflexive relation,
$p$, $p_1$ and $p_2$ are sets of states,
and,
\begin{align}
  & \EstablishExpr{p}{r}{b_1}{\True}{p_1}  \labelprop{post-conj1} \\
  & \EstablishExpr{p}{r}{b_2}{\True}{p_2}  \labelprop{post-conj2} 
\end{align}
then,
$\EstablishExpr{p}{r}{\Binary{b_1}{\land}{b_2}}{\True}{p_1 \inter p_2}$.
\end{lemmax}

\begin{proof}
The proof applies \reflem{post-binary} with the two functions 
$P_1 \defs (\lambda k \spot p_1)$ and $P_2 \defs (\lambda k \spot p_2)$, that simply ignore their parameters.
The assumption of \reflem*{post-binary} \refprop{post-binary-promote} with $\True$ for $\Ex{k}$ requires $\Ex{\True = k_1 \land k_2}$,
which implies both $\Ex{k_1}$ and $\Ex{k_2}$ are $\Ex{\True}$,
and hence the first two assumptions \refprop{post-binary1} and \refprop{post-binary2} of \reflem*{post-binary} 
hold by \refprop{post-conj1} and \refprop{post-conj2}.
Finally, $P_1\,\Ex{\True} \inter P_2\,\Ex{\True} \subseteq p_1 \inter p_2$, holds because, 
$P_1\,\Ex{\True} \inter P_2\,\Ex{\True}$, reduces to, $p_1 \inter p_2$.
\end{proof}

\reflem{post-disjoin} establishes a postcondition, $p_1 \union p_2$, under the same assumptions as \reflem{post-conjoin}.
\begin{lemmax}[post-disjoin] 
If $\Ex{b_1}$ and $\Ex{b_2}$ are boolean expressions,
$r$ is a reflexive relation,
$p$, $p_1$ and $p_2$ are sets of states, and
\begin{align}
  & \EstablishExpr{p}{r}{b_1}{\True}{p_1}  \labelprop{post-disj1} \\
  & \EstablishExpr{p}{r}{b_2}{\True}{p_2}  \labelprop{post-disj2} 
\end{align}
then,
$\EstablishExpr{p}{r}{\Binary{b_1}{\lor}{b_2}}{\True}{p_1 \union p_2}$.
\end{lemmax}

\begin{proof}
The proof applies \reflem{post-binary} with functions 
$P_1 = (\lambda \Ex{k} \spot \If \Ex{k = \True} \Then p_1 \Else \Sigma)$ and 
$P_2 = (\lambda \Ex{k} \spot \If \Ex{k = \True} \Then p_2 \Else \Sigma)$.
For $\Ex{k_1}$ and $\Ex{k_2}$ both $\Ex{\True}$, 
the first two assumptions \refprop{post-binary1} and \refprop{post-binary2} for \reflem*{post-binary} 
reduce to our assumptions \refprop{post-disj1} and \refprop{post-disj2},
and for $\Ex{k_1}$ and $\Ex{k_2}$ as $\Ex{\False}$, $P_1\,\Ex{k_1}$ and $P_2\,\Ex{k_2}$ both reduce to $\Pre{\Sigma}$, 
and hence the first two assumptions \refprop{post-binary1} and \refprop{post-binary2} for \reflem*{post-binary} trivially hold in this case.
For $\Ex{k}$ true, either $\Ex{k_1}$ or $\Ex{k_2}$ is $\Ex{\True}$, and $P_1\,\Ex{k_1} \inter P_2\,\Ex{k_2}$ reduces to either $p_1 \inter \Sigma$ or $\Sigma \inter p_2$ or $p_1 \inter p_2$,
all of which are contained in $p_1 \union p_2$, and hence the final assumption \refprop{post-binary-promote} holds for the application of \reflem*{post-binary}.
\end{proof}
\reflem*{post-conjoin} and \reflem*{post-disjoin} may be recursively applied to break down arbitrarily nested conjunctions and disjunctions 
until the ``leaf'' expressions are base cases, at which point \reflem{post-constant} or \reflem{post-variable} can be applied.
Any logical negations can be pushed down to ``leaf'' expressions that are base cases
using laws (\refprop*{expr-not}--\refprop*{de-morgan2})
and then \reflem{post-unary} applied to handle the logical negations.

Note that for the base cases, \reflem{post-constant} and \reflem{post-variable},
both the precondition and the postcondition are required to be stable under $r$
but the lemmas in this section do not explicitly assume these stability requirements.
Of course these lemmas do have assumptions about their sub-components 
and for the base cases the stability conditions are required for the leaf expressions.
In practice that means when we apply these lemmas $p$ and $P\,\Ex{k}$ are typically stable under $r$,
noting that if both $p_1$ and $p_2$ are stable under $r$, so are $p_1 \inter p_2$ and $p_1 \union p_2$.

\subsection{Introducing expression evaluations, including guards}\labelsect{intro-expr}

The following lemmas are used to introduce an expression evaluation, along with its established postcondition.
These laws are used to prove laws for introducing assignment, conditional and loop commands in \refsect{commands}.
The laws in this subsection apply to both expressions and l-value expressions;
we just present the expression versions here.
\begin{lemmax}[idle-expr-with-post]
If 
$\EstablishExpr{p}{r}{e}{k}{P\,\Ex{k}}$,
\begin{align*}
  \rely{r} \together \Pre{p} \Seq \Idle \refsto \rely{r} \together \Expr{e}{k} \Seq \Pre{P\,\Ex{k}} .
\end{align*}
\end{lemmax}

\begin{proof}
The proof uses \refprop{idle-to-expr} to introduce the expression evaluation $\Expr{e}{k}$ 
and then applies \refprop{triple-with-rely} using the assumption:
$\rely{r} \together \Pre{p} \Seq \Idle 
 \refsto \rely{r} \together \Pre{p} \Seq \Expr{e}{k}
 \refsto \rely{r} \together \Expr{e}{k} \Seq \Pre{P\,\Ex{k}}$. 
\end{proof}

The following lemma allows an expression evaluation to be introduced before a command $c$ 
provided $c$ tolerates interference satisfying a rely condition $r$ before the start of execution of $c$ from initial states satisfying $p$;
this is expressed formally by property \refprop{tolerates-r}.
\begin{lemmax}[introduce-expr-with-post]
If both
\begin{align}
  & \EstablishExpr{p}{r}{e}{k}{P\,\Ex{k}}  \labelprop{intro-establish-fk} \\
  & \rely{r} \together \Pre{p} \Seq c \refsto \rely{r} \together \Pre{p} \Seq \Idle \Seq c \labelprop{tolerates-r}
\end{align}
then,
\(
  \rely{r} \together \Pre{p} \Seq c \refsto \Expr{e}{k} \Seq \Pre{P\,\Ex{k}} \Seq (\rely{r} \together c) .
\)
\end{lemmax}

\begin{proof}
\begin{align*}&
  \rely{r} \together \Pre{p} \Seq c 
 \Refsto*[by \refprop{tolerates-r} and distribute the rely \refprop{rely-distrib-seq}] 
  (\rely{r} \together \Pre{p} \Seq \Idle) \Seq (\rely{r} \together c)
 \Refsto*[by \reflem{idle-expr-with-post} using \refprop{intro-establish-fk}]
  \Expr{e}{k} \Seq \Pre{P\,\Ex{k}} \Seq (\rely{r} \together c)
 \qedhere
\end{align*}
\end{proof}

A specification command \refdef{spec} with a relational postcondition $q$ 
needs to be able to tolerate interference satisfying the rely condition $r$ both before and after executing.
\begin{definitionx}[tolerates]
A relation $q$ \emph{tolerates} interference $r$ from precondition $p$ if $p$ is stable under $r$ and both,
\begin{align}
  p \dres (r \semi q) \subseteq q  
  ~~~~~\mbox{and}~~~~~
  p \dres (q \semi r) \subseteq q , 
\end{align}
where $\semi$ is (forward) composition of binary relations,
that is, $(\sigma,\sigma') \in (r \semi q)$ if and only if $(\exists \sigma_1 \spot (\sigma,\sigma_1) \in r \land (\sigma_1,\sigma') \in q)$.
\end{definitionx}
If $q$ tolerates $r$ from $p$, it is shown in \cite{hayes2021deriving} that
$p \dres (\Finrel{r} \semi q \semi \Finrel{r}) \subseteq q$,
where $\Finrel{r}$ is the reflexive transitive closure of the relation $r$.
That is, $q$ tolerates zero or more $r$ transitions both before and after it,
and thus one can introduce $\Idle$ commands before and after a specification command with postcondition $q$
(see \cite{hayes2021deriving} for a proof of this property).
\begin{align}
  \rely{r} \together \Pre{p} \Seq \Spec{}{}{q} & = \rely{r} \together \Pre{p} \Seq \Idle \Seq \Spec{}{}{q} \Seq \Idle & \mbox{if $q$ tolerates $r$ from $p$} \labelprop{spec-tolerates}
\end{align}

\subsection{Ensuring expressions are well defined and well typed}\labelsect{well-defined}

Our programming language is not statically typed and hence we need to establish that expression evaluations are correctly typed,
for example, that the guard, $\Ex{b}$, of a conditional \refdef{conditional} or while loop \refdef{while} evaluates to a boolean value,
that is, for all $\Ex{k}$ it satisfies the following triple,
\begin{align}
  \EstablishExpr{p}{r}{b}{k}{\Set{k \in \bool}}   \labelprop{cond-type-bool}
\end{align}
When applying the rules for introducing conditionals and while loops below,
they need to rule out their aborting behaviours by ensuring the guard evaluates to a boolean.
The following lemma generalises this from type boolean to an arbitrary type $\Ex{S}$.
\begin{lemmax}[expr-non-type]
For an expression $\Ex{e}$ if,
\begin{align}
  \rely{r} \together \Pre{p} \Seq c  \refsto \rely{r} \together \Pre{p} \Seq \Idle \Seq c \labelprop{c-tolerates-r} \\
  \forall \Ex{k} \spot  \EstablishExpr{p}{r}{e}{k}{\Set{k \in S}}   \labelprop{cond-type} 
\end{align}
then, $\rely{r} \together \Pre{p} \Seq c \refsto \Nondet_{\Ex{k}}^{\Ex{k \notin S}} \Expr{e}{k} \Seq \Abort$.
\end{lemmax}

\begin{proof}
\begin{align*}&
  \rely{r} \together \Pre{p} \Seq c
 \Refsto*[apply \refprop{c-tolerates-r} and introduce non-deterministic choice over $k \notin S$]
  \Nondet_{\Ex{k}}^{\Ex{k \notin S}} \rely{r} \together \Pre{p} \Seq \Idle \Seq c 
 \Refsto*[apply \reflem{idle-expr-with-post} using \refprop{cond-type} and remove rely \refprop{rely-remove}]
  \Nondet_{\Ex{k}}^{\Ex{k \notin S}} \Expr{e}{k} \Seq \Pre{\Set{k \in S}} \Seq c 
 \Equals*[for $k \notin S$, we have $\Pre{\Set{k \in S}} = \Pre{\emptyset} = \Abort$ and $\Abort$ annihilates from the left]
  \Nondet_{\Ex{k}}^{\Ex{k \notin S}} \Expr{e}{k} \Seq \Abort 
 \qedhere
\end{align*}
\end{proof}

\section{Refinement laws for commands using expressions}\labelsect{commands}

This section develops laws for reasoning about conditionals \refsect{conditional}, and assignments \refsect{assign}.

\subsection{Conditional commands}\labelsect{conditional}

A conditional command either evaluates its boolean guard $\Ex{b}$ to $\Ex{\True}$ and executes $c$, 
or to $\Ex{\False}$ and executes $d$ \refdef{conditional}.
Because our language is not statically typed, the semantics needs to deal with guards that evaluate to a non-boolean,
in which case conditionals (and loops) are defined to abort.
The law for conditionals below includes an assumption \refprop{cond-bool}
that avoids the case when $\Ex{b}$ evaluates to a non-boolean.
\begin{align}
  \If b \Then c \Else d \Fi & \defs \Expr{b}{\True} \Seq c \nondet \Expr{b}{\False} \Seq d \nondet \Nondet_{\Ex{k}}^{\Ex{k \not\in \bool}} \Expr{b}{k} \Seq \Abort \labeldef{conditional} 
\end{align}
\begin{minipage}{0.7\textwidth}
As suggested at the start of \refsect{motivation} for property  \refprop{if-pt-pf},
a conditional command in the context of interference satisfying a rely condition $r$, 
makes use of assumptions that its guard evaluations establish postconditions 
that can be shown using the lemmas in \refsect{complex-expressions}.
Separate assumptions are required for the case of the guard $\Ex{b}$ evaluating to $\Ex{\True}$ \refprop{cond-infer-true}
and $\Ex{b}$ evaluating to $\Ex{\False}$ \refprop{cond-infer-false}.
The following theorem formalises property \refprop{if-pt-pf}.
\end{minipage}%
\begin{minipage}{0.3\textwidth}
\begin{center} 
\begin{picture}(0,0)%
\includegraphics{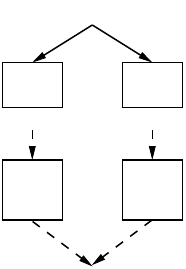}%
\end{picture}%
\setlength{\unitlength}{3158sp}%
\begin{picture}(1844,2683)(5679,-4598)
\put(6601,-3361){\makebox(0,0)[b]{\smash{\fontsize{10}{12}\usefont{T1}{ptm}{m}{n}{\color[rgb]{0,0,0}$\Finrel{r}$}%
}}}
\put(6601,-2086){\makebox(0,0)[b]{\smash{\fontsize{10}{12}\usefont{T1}{phv}{m}{n}{\color[rgb]{0,0,0}$p$}%
}}}
\put(7201,-3736){\makebox(0,0)[b]{\smash{\fontsize{10}{12}\usefont{T1}{phv}{m}{n}{\color[rgb]{0,0,0}else}%
}}}
\put(6151,-4411){\makebox(0,0)[rb]{\smash{\fontsize{10}{12}\usefont{T1}{phv}{m}{n}{\color[rgb]{0,0,0}$\Finrel{r}$}%
}}}
\put(7051,-4411){\makebox(0,0)[lb]{\smash{\fontsize{10}{12}\usefont{T1}{phv}{m}{n}{\color[rgb]{0,0,0}$\Finrel{r}$}%
}}}
\put(6001,-3961){\makebox(0,0)[b]{\smash{\fontsize{10}{12}\usefont{T1}{phv}{m}{n}{\color[rgb]{0,0,0}$c$}%
}}}
\put(7201,-3961){\makebox(0,0)[b]{\smash{\fontsize{10}{12}\usefont{T1}{phv}{m}{n}{\color[rgb]{0,0,0}$c$}%
}}}
\put(6001,-3736){\makebox(0,0)[b]{\smash{\fontsize{10}{12}\usefont{T1}{phv}{m}{n}{\color[rgb]{0,0,0}then}%
}}}
\put(6001,-2836){\makebox(0,0)[b]{\smash{\fontsize{10}{12}\usefont{T1}{ptm}{m}{n}{\color[rgb]{0,0,0}$\Expr{b}{\True}$}%
}}}
\put(7201,-2836){\makebox(0,0)[b]{\smash{\fontsize{10}{12}\usefont{T1}{ptm}{m}{n}{\color[rgb]{0,0,0}$\Expr{b}{\False}$}%
}}}
\put(6001,-3136){\makebox(0,0)[b]{\smash{\fontsize{10}{12}\usefont{T1}{phv}{m}{n}{\color[rgb]{0,0,0}$p_t$}%
}}}
\put(7201,-3136){\makebox(0,0)[b]{\smash{\fontsize{10}{12}\usefont{T1}{phv}{m}{n}{\color[rgb]{0,0,0}$p_f$}%
}}}
\put(6601,-3886){\makebox(0,0)[b]{\smash{\fontsize{10}{12}\usefont{T1}{ptm}{m}{n}{\color[rgb]{0,0,0}$\Finrel{r}$}%
}}}
\put(6601,-2836){\makebox(0,0)[b]{\smash{\fontsize{10}{12}\usefont{T1}{phv}{m}{n}{\color[rgb]{0,0,0}$\Finrel{r}$}%
}}}
\end{picture}%
 
\end{center}
\end{minipage}

\begin{theoremx}[conditional-inference]
For a relation $r$,
sets of states $p$, $p_t$ and $p_f$,
and
commands $c$ and $d$,
if
\begin{align}
  & \EstablishExpr{p}{r}{b}{\True}{p_t}   \labelprop{cond-infer-true} \\
  & \EstablishExpr{p}{r}{b}{\False}{p_f}   \labelprop{cond-infer-false} 
\end{align}
then,
$\rely{r} \together \Pre{p} \Seq \If b \Then c \Else d \Fi = \rely{r} \together \Pre{p} \Seq \If b \Then \Pre{p_t} \Seq c \Else \Pre{p_f} \Seq d \Fi$.
\end{theoremx}

\begin{proof}
\begin{align*}&
  \rely{r} \together \Pre{p} \Seq \If b \Then c \Else d \Fi
 \Equals*[unfold the definition of a condition \refdef{conditional} and distributing the precondition $p$]
  \rely{r} \together \Pre{p} \Seq \Expr{b}{\True} \Seq c \nondet \Pre{p} \Seq \Expr{b}{\False} \Seq d \nondet \Pre{p} \Seq \Nondet_{\Ex{k}}^{\Ex{k \not\in \bool}} \Expr{b}{k} \Seq \Abort 
 \Equals*[by assumptions \refprop{cond-infer-true} and \refprop{cond-infer-false}]
  \rely{r} \together \Pre{p} \Seq \Expr{b}{\True} \Seq \Pre{p_t} \Seq c \nondet \Pre{p} \Seq \Expr{b}{\False} \Seq\Pre{p_f} \Seq d \nondet \Pre{p} \Seq \Nondet_{\Ex{k}}^{\Ex{k \not\in \bool}} \Expr{b}{k} \Seq \Abort 
 \Equals*[undistribute the precondition $p$ and fold the definition of the conditional \refdef{conditional}]
  \rely{r} \together \Pre{p} \Seq \If b \Then \Pre{p_t} \Seq c \Else \Pre{p_f} \Seq d \Fi
 \qedhere
\end{align*}
\end{proof}

For the law for introducing a conditional command, 
as well as the assumptions about the guard $\Ex{b}$ evaluating to true establishing $p_t$ \refprop{cond-true} 
and to false establishing $p_f$ \refprop{cond-false},
a third assumption \refprop{cond-bool} ensures $\Ex{b}$ evaluates to a boolean value.
In addition, the command being refined, $c$, must tolerate interference satisfying $r$ before it begins \refprop{c-tolerates-idle}
in order to allow for the evaluation of the guard occurring before $c$ begins.

\begin{lawx}[intro-conditional]
Given a boolean expression $\Ex{b}$, 
a relation $r$, 
sets of states $p$, $p_t$ and $p_f$,
and a command $c$,
if
\begin{align}
  & \EstablishExpr{p}{r}{b}{\True}{p_t}   \labelprop{cond-true} \\
  & \EstablishExpr{p}{r}{b}{\False}{p_f}   \labelprop{cond-false} \\
  \forall k \spot & \EstablishExpr{p}{r}{b}{k}{\Set{k \in \bool}}   \labelprop{cond-bool} \\
  \rely{r} \together \Pre{p} \Seq c & \refsto \rely{r} \together \Pre{p} \Seq \Idle \Seq c \labelprop{c-tolerates-idle}
\end{align}
then
\(
  \rely{r} \together \Pre{p} \Seq c \refsto \If \Ex{b} \Then \rely{r} \together \Pre{p_t} \Seq c \Else \rely{r} \together \Pre{p_f} \Seq c \Fi .
\)
\end{lawx}

\begin{proof}
The proof uses \reflem{introduce-expr-with-post} to introduce $\Expr{b}{\True}$ and $\Expr{b}{\False}$
and \reflem{expr-non-type} to handle the non-boolean case.
\begin{align*}&
  \rely{r} \together \Pre{p} \Seq c 
 \Equals*[non-deterministic choice is idempotent]
  \rely{r} \together \Pre{p} \Seq c \nondet \rely{r} \together \Pre{p} \Seq c \nondet \rely{r} \together \Pre{p} \Seq c 
 \Refsto*[by \reflem*{introduce-expr-with-post} twice using \refprop{cond-true}, \refprop{cond-false} and \refprop{c-tolerates-idle} and \reflem*{expr-non-type} using \refprop{cond-bool}]
  \Expr{b}{\True} \Seq (\rely{r} \together \Pre{p_t} \Seq c) \nondet 
  \Expr{b}{\False} \Seq (\rely{r} \together \Pre{p_f} \Seq c) \nondet 
  \Nondet_{\Ex{k}}^{\Ex{k \not\in \bool}} \Expr{b}{k} \Seq \Abort
 \Equals*[definition of conditional \refdef{conditional}]
  \If \Ex{b} \Then \rely{r} \together \Pre{p_t} \Seq c \Else \rely{r} \together \Pre{p_f} \Seq c \Fi
 \qedhere
\end{align*}
\end{proof}

If a relation $q$ tolerates $r$ from $p$,
by \refprop{spec-tolerates} the specification command $\Post{q}$ satisfies the proof obligation \refprop{c-tolerates-idle} of \reflaw*{intro-conditional}
giving the following corollary.
\begin{corx}[intro-conditional-spec]
\reflaw{intro-conditional} holds if $c$ is a specification of the form $\Post{q}$ such that $q$ tolerates $r$ from $p$.
\end{corx}

Note that guarantees distribute over non-deterministic choice and sequential composition, 
and expression evaluation and $\Idle$ satisfy a reflexive guarantee,
and hence a reflexive guarantee can be distributed into a conditional command
(see \cite{hayes2021deriving} for a detailed proof).

A C/Java style conditional expression, $\Ex{\Cond{b}{e_1}{e_2}}$, is straightforward to define in terms of a conditional command \refdef{cond-expr},
and conditional ``and'' and ``or'' operations can be defined in terms of it \refdef{cond-and}--\refdef{cond-or}.
We use the syntax $\Cand$ and $\Cor$ for the operations to avoid confusion of $\parallel$ for parallel composition 
with the C-style $||$ for conditional or.
\begin{align}
  \Expr{\Cond{b}{e_1}{e_2}}{k} & \defs \If \Ex{b} \Then \Expr{e_1}{k} \Else \Expr{e_2}{k} \Fi \labeldef{cond-expr} \\
  \Ex{b_1 \Cand b_2} & \defs (\Ex{\Cond{b_1}{b_2}{\False}}) \labeldef{cond-and} \\
  \Ex{b_1 \Cor b_2} & \defs (\Ex{\Cond{b_1}{\True}{b_2}}) \labeldef{cond-or}
\end{align}
Reasoning about these forms of expression can then be performed using the law for conditionals.

\subsection{While loops}\labelsect{loop}

A while loop executes $c$ while $\Ex{b}$ evaluates to $\Ex{\True}$ and then terminates if $\Ex{b}$ evaluates to $\Ex{\False}$
or aborts if $\Ex{b}$ evaluates to a non-boolean  \refdef{while}.
Note that because the definition is in the form of a greatest fixed point, 
it allows possibly infinite iteration of $c$, 
so that our definition of a while loop allows one to address loop termination.
\begin{align}
  \While b \Do c \Od & \defs \nu x .~ \If b \Then c \Seq x \Else \Nil \Fi \labeldef{while}
\end{align}
A law similar to \Theorem{conditional-inference} can be used to justify property \refprop{while-inference} for while loops
but reasoning about while loops requires laws for reasoning about greatest fixed points (recursion)
and showing termination of while loops requires laws involving well-founded relations,
both of which are beyond the scope of the current paper 
(see \cite{2026RecursionWhileLoops} for a treatment of recursion and while loops in our theory).

\subsection{Introducing an assignment command}\labelsect{assign}

An assignment command, $\Assignment{lve}{e}$, 
is non-deterministic due to possible interference from concurrent threads modifying the values of the variables
used within $\LEx{lve}$ or $\Ex{e}$.
Recall that, $set\,\LEx{lv}\,\Ex{k}\,\sigma$, 
stands for the state consisting of state $\sigma$ updated so that l-value $\LEx{lv}$ takes on value $\Ex{k}$.
An assignment evaluates its l-value expression $\LEx{lve}$ to an l-value $\LEx{lv}$ 
and its expression $\Ex{e}$ to some value $\Ex{k}$, 
which is then atomically assigned to $\LEx{lv}$ \refdef{assignment}.
An optional command \refdef{opt} is used to allow an assignment of the form $\Assignment{lve}{\Deref{lve}}$ to do nothing and
the $\Idle$ at the end allows for stuttering steps in the implementation.
The relation $update\,\LEx{lv}\,\Ex{k}$ updates the state so that $\LEx{lv}$ takes the value $\Ex{k}$.
\begin{align}
  update\,\LEx{lv}\,\Ex{k} & \defs \{(\sigma,\sigma') \spot \sigma' = set\,\LEx{lv}\,\Ex{k}\,\sigma\} \labeldef{update} \\
  \Assignment{lve}{e} & \defs \Nondet_{\LEx{lv},\Ex{k}} (\LExpr{lve}{lv} \parallel \Expr{e}{k}) \Seq \opt(update\,\LEx{lv}\,\Ex{k}) \Seq \Idle \labeldef{assignment} 
\end{align}
For an assignment command, $\Assignment{v}{e}$, if $\Ex{e}$ evaluates to a value $\Ex{k}$,
one cannot assume $\EqEval{e}{k}$ still holds when $\LEx{v}$ is updated by the assignment.
For example, for $\Assignment{v}{\Deref{u}}$, if under inference, $r$, such that $\Variable{v}$ is not increased and $\Variable{u}$ is not decreased,
one cannot assume $\Set{\Deref{v} = \Deref{u}}$ after the assignment, only that $\Set{\Deref{v} \leq \Deref{u}}$.
See Example \ref{ex-assign-v-w} below for a formal treatment of this example.
Similarly, if the left side of an assignment is an indexed array element, $\Array{lve}{e}$
and the index expression $\Ex{e}$ is subject to interference then after executing $\LExpr{\Array{lve}{e}}{lv}$
one cannot deduce that $\Array{lve}{e}$ is equal to $\LEx{lv}$.
The laws in \refsect{complex-expressions} can be used to establish assumptions \refprop{assign-lve} and \refprop{assign-expr} 
for the law for introducing an assignment command.
The state update in the assignment must satisfy both the guarantee $g$ and the postcondition $q$ \refprop{assign-opt}.
\begin{lawx}[rely-guar-assign]
Given an expression $\Ex{e}$,
an l-value expression $\LEx{lve}$ that has atomic access,
reflexive relations $r$ and $g$,
a set of states $p$,
a function $P_1$ from l-values to sets of states,
a function $P_2$ from values to sets of states,
and a relation $q$ that tolerates $r$ from $p$,
if,
\begin{align}
  & \forall \LEx{lv} \spot \EstablishLExpr{p}{r}{lve}{lv}{P_1\,\LEx{lv}} \labelprop{assign-lve} \\
  & \forall \Ex{k} \spot \EstablishExpr{p}{r}{e}{k}{P_2\,\Ex{k}} \labelprop{assign-expr} \\
  & \forall lv\,k \spot(P_1\,\LEx{lv} \inter P_2\,\Ex{k}) \dres update\,\LEx{lv}\,\Ex{k} \subseteq g \inter q  \labelprop{assign-opt}
\end{align}
then,
\(
  \guar{g} \together \rely{r} \together \Pre{p} \Seq \Spec{}{}{q} \refsto \Assignment{v}{e} .
\)
\end{lawx}
An additional proviso to handle an undefined expression evaluating to the undefined value $\UndefinedValue$ is,
\begin{align*}
  \EstablishExpr{p}{r}{e}{k}{\Set{k \neq \UndefinedValue}} 
\end{align*}
In practice one usually requires a stronger typing invariant, for example, $\Ex{k \in int}$,
which rules out $\Ex{k = \UndefinedValue}$, and hence subsumes checking for undefined expressions.
\begin{proof}
By Parallel Hoare inference rule \refprop{hoare-parallel} and assumptions \refprop{assign-lve} and \refprop{assign-expr} we can infer,
\begin{align}
\forall \LEx{lv}, \Ex{k} \spot
\Triple{p}{&\rely{r} \together (\LExpr{lve}{lv} \parallel \Expr{e}{k})}{P_1\,\LEx{lv} \inter P_2\,\Ex{k}} .
\labelprop{rely-guar-assign-triple}
\end{align}
By applying \refprop{spec-to-opt} using assumption \refprop{assign-opt} that,
\begin{align}
  \forall \LEx{lv}\,\Ex{k} \spot \Pre{P_1\,\LEx{lv} \inter P_2\,\Ex{k}} \Seq (\guar{g} \together \Spec{}{}{q}) \refsto \opt(update\,\LEx{lv}\,\Ex{k}) .\labelprop{assign-guar}
\end{align}
Now we can show the law holds as follows.
\begin{align*}&
  \guar{g} \together \rely{r} \together \Pre{p} \Seq \Post{q}\\
\Equals*[$q$ tolerates $r$ from $p$ \refprop{spec-tolerates}]
  \guar{g} \together \rely{r} \together \Pre{p} \Seq \Idle \Seq \Post{q} \Seq \Idle
 \Equals*[distributing the guarantee by \refprop{guar-distrib-seq} and $\Idle$ satisfies a reflexive guarantee \refprop{idle-guar-reflexive}]
  \rely{r} \together \Pre{p} \Seq \Idle \Seq (\guar{g} \together \Post{q}) \Seq \Idle
 \Refsto*[introduce choice and $\Idle = \Idle \parallel \Idle$ and expression evaluations refine $\Idle$ \refprop{idle-to-expr}]
  \Nondet_{\LEx{lv}\,\Ex{k}} \spot\rely{r} \together \Pre{p} \Seq (\LExpr{lve}{lv} \parallel \Expr{e}{k}) \Seq (\guar{g} \together \Post{q}) \Seq \Idle
 \Refsto*[apply \refprop{triple-with-rely} using \refprop{rely-guar-assign-triple} and remove the rely \refprop{rely-remove}]
  \Nondet_{\LEx{lv}\,\Ex{k}} \spot (\LExpr{lve}{lv} \parallel \Expr{e}{k}) \Seq \Pre{P_1\,\LEx{lv} \inter P_2\,\Ex{k}} \Seq (\guar{g} \together \Post{q}) \Seq \Idle
 \Refsto*[apply \refprop{assign-guar}]
  \Nondet_{\LEx{lv}\,\Ex{k}} \spot (\LExpr{lve}{lv} \parallel \Expr{e}{k}) \Seq \opt(update\,\LEx{lv}\,\Ex{k})  \Seq \Idle
 \Equals*[definition of an assignment command \refdef{assignment}] 
  \Assignment{lve}{e}
 \qedhere
\end{align*}
\end{proof}

\begin{example}\label{ex-assign-v-w}
\reflaw{rely-guar-assign} can be used to show the following refinement.
\begin{align}
  \Rguar{\Deref{u}' = \Deref{u}} \together \Rrely{\Deref{v}' \leq \Deref{v} \land \Deref{u} \leq \Deref{u}'} \together \Pre{\Set{\True}} \Seq \RPost{\Deref{v}' \leq \Deref{u}'} \refsto \Assignment{v}{\Deref{u}}
\end{align}
The postcondition $\Rel{\Deref{v}' \leq \Deref{u}'}$ tolerates the rely condition because $\Set{\Deref{v} \leq \Deref{u}}$ is stable under the rely.
Proof obligation \refprop{assign-expr} holds by \reflem{post-variable} if we take $P\,\Ex{k}$ to be $\Set{k \leq \Deref{u}}$ 
because $\Ex{k = \Deref{u} \implies k \leq \Deref{u}}$, and $\Set{k \leq \Deref{u}}$ is stable under the rely.
If $\LEx{lv}$ is $\Variable{v}$, proof obligation \refprop{assign-opt} corresponds to
$\Set{k \leq \Deref{u}}  \inter \{(\sigma,\sigma') \spot \sigma' = set\,\LEx{v}\,\Ex{k}\,\sigma\} \subseteq \Rel{\Deref{u}' = \Deref{u} \land \Deref{v}' \leq \Deref{u}'}$,
which holds because $set\,\LEx{v}\,\Ex{k}$ sets $\LEx{v}$ to $\Ex{k}$, which is is no greater than $\Variable{u}$, and does not change $\Variable{u}$.
\end{example}

\begin{RelatedWork}
Note that earlier approaches with the syntactic single critical reference constraint cannot handle Example~\ref{ex-assign-v-w}
because both $\Variable{v}$ and $\Variable{u}$ can be modified by interference.
\end{RelatedWork}

\section{Example: Fischer-Galler representation of an equivalence relation}\labelsect{equiv-example}

Consider a data structure representing an equivalence relation with two operations that can execute concurrently: 
$test(x,y)$ that tests whether elements $x$ and $y$ are equivalent,
and 
$equate(x,y)$ that equates $x$ with $y$, 
forming its reflexive transitive closure to maintain the property that it is an equivalence relation.
The equivalence relation initially consists of the identity relation, so that each element is only related to itself.
The implementation makes use of a Fischer-Galler forest-of-trees data structure, $f$,
where two elements are equivalent in $f$ if they are in the same tree within the forest;
being in the ``same tree'' is determined by comparing their roots \cite{1964GallerFischer}.
(This example is tackled in~\cite{Jones81d} and~\cite{ColletteJones00a},
where both make use of a semantics that assumes expression evaluation and assignment commands are atomic;
in the former the focus is on a concurrent clean-up operation.)

\subsection{The test operation}\labelsect{equiv-test}

Allowing concurrent $equate$ operations requires that the specification of the $test$ operation needs to be non-deterministic 
because of potential interference:
if $\Variable{x}$ and $\Variable{y}$ are equivalent in the initial forest $\Deref{f}$, when $test$ is called, it must return $\Ex{\True}$, and
if $\Variable{x}$ and $\Variable{y}$ are not equivalent in the final forest $\Deref{f'}$, when $test$ exits, it must return $\Ex{\False}$,
but if $\Variable{x}$ and $\Variable{y}$ are not equivalent in $\Ex{f}$ but are equivalent in $\Ex{f'}$
(i.e they have become equivalent via a concurrent $equate$ during the execution of $test$)
then $test$ may return either $\Ex{\True}$ or $\Ex{\False}$.
This gives the following post-condition for $test$, in which $\Deref{t}'$ is the value of the result of a call,
and for any elements $v$ and $u$, $v \equiv_f u$ means that $v$ and $u$ are equivalent in the forest $f$.
\begin{align}
  ((\Deref{x} \equiv_{\Deref{f}} \Deref{y}) \implies \Deref{t}') \land (\Deref{t}' \implies (\Deref{x} \equiv_{\Deref{f}'} \Deref{y})) \labelprop{test-post}
\end{align}
The function $test$ in \reffig{test} determines whether or not $\Variable{x}$ and $\Variable{y}$ are equivalent
by checking whether the roots of their trees are the same.
To find the root of the tree containing element $v$, parent links in the tree are followed until the root is found,
where the parent link for a root $rv$ points to itself 
(i.e.\ $rv \in roots(f)$ iff $rv = f[rv]$).
Interference can occur during the while loop, including during evaluation of its guard.
\begin{figure}[ht]
\begin{align*}
  \begin{array}{l}
    \Variable{t} : \bool \leftarrow test(\Variable{x}, \Variable{y} : X) \\
     \begin{array}{l}
      \Var \Variable{rx},\Variable{ry} : X; \\
      (\Assignment{rx}{\Deref{x}} \parallel \Assignment{ry}{\Deref{y}}); \\
      \{(\Deref{x} \equiv_{\Deref{f}} \Deref{rx}) \land (\Deref{y} \equiv_{\Deref{f}} \Deref{ry})\} \\
      \While \Deref{rx} \neq \Deref{(f[\Ex{\Deref{rx}}])} \lor \Deref{ry} \neq \Deref{(f[\Ex{\Deref{ry}}])} \Do {}\\
       \begin{array}{l}
        (\Spec{rx}{}{\Rel{\Deref{rx}' \equiv_{\Deref{f}} \Deref{x} \land \Deref{rx}' \in roots(\Deref{f})}} \parallel 
         \Spec{ry}{}{\Rel{\Deref{ry}' \equiv_{\Deref{f}} \Deref{y} \land \Deref{ry}' \in roots(\Deref{f})}})
       \end{array} \\
      \Od; \\
      \Assignment{t}{(\Deref{rx} = \Deref{ry})}
     \end{array}
  \end{array}
\end{align*}
\caption{Operation to test equivalence with multiple references to $f$ in its guard}\labelfig{test}
\Description{Code for the test operation.}
\end{figure}

Once $\Variable{rx}$ and $\Variable{ry}$ are initially assigned, the code for $test$ maintains the invariant,
\begin{align}
  (\Deref{x} \equiv_{\Deref{f}} \Deref{rx}) \land (\Deref{y} \equiv_{\Deref{f}} \Deref{ry}) . \labelprop{test-inv}
\end{align}
For two forests $f_1$ and $f_2$, we use $f_1 \subseteq f_2$ to mean the equivalence relation represented by $f_1$ is contained in that of $f_2$, that is,
\begin{align}
  f_1 \subseteq f_2 & \defs \forall v\,u \spot (v \equiv_{f_1} u) \implies (v \equiv_{f_2} u) . \labeldef{f-grows}
\end{align}
There is no operation to delete equivalences within $\Variable{f}$, 
so once a pair of elements $v$ and $u$ are equivalent, they remain so.
In the concurrent case, interference from parallel $equate$ operations (not given in detail here) may grow $\Variable{f}$
and hence the rely condition of $test$ implies $\Rel{\Deref{f} \subseteq \Deref{f}'}$.
Note that postcondition \refprop{test-post} tolerates interference represented by this rely condition.
An operation to equate $v$ and $u$, may modify the forest so that the parent of the root of $v$ is set to equal the root of $u$ 
(or vice versa),
so that the root of $v$ is now the same as the root of $u$.
The $equate$ operation may change the forest so that an element that was a root is no longer a root ---
but note that an element that is not a root can never become a root.
Combining this with $\Rel{\Deref{f} \subseteq \Deref{f}'}$ gives the rely condition $r$ for $test$.
\begin{align}
  r \defs \Rel{\Deref{f} \subseteq \Deref{f}' \land roots(\Deref{f}') \subseteq roots(\Deref{f})} . \labeldef{test-rely}
\end{align}
\begin{RelatedWork}
Both the check whether $\Variable{rx}$ is a root and whether $\Variable{ry}$ is a root in the loop guard of $test$
may be affected by interference from concurrent equates
and hence $test$ in \reffig{test} does not meet the syntactic single critical reference constraint 
required in the previous approaches
\cite{Jones81d,Jones83a,Jones83b,Stolen90,stolen1991method,XuRoeverHe97,Dingel02,PrensaNieto03,Wickerson10-TR,DBLP:conf/esop/WickersonDP10,SchellhornTEPR14,Sanan21}.
Of course, local variables can be introduced to capture the result of each of 
$\Deref{rx} \neq \Deref{(\Array{f}{\Deref{rx}})}$ and $\Deref{ry} \neq \Deref{(\Array{f}{\Deref{ry}})}$,
but that complicates the structure of the program.
It also complicates the reasoning to show it is correct
because the reasoning requires both additional inference steps and 
a more complex invariant to accommodate the purpose for the extra local variables.
And the ``single critical reference'' constraint of earlier approaches is not formally provable in their theories.
\end{RelatedWork}

\subsection{Reasoning about the guard within test}\labelsect{test-guard}

To illustrate our approach to reasoning about the loop guards,
we derive the assertions for the precondition of the body of the loop and the exit condition for the loop within $test$.
For this section we augment the rely condition from the specification \refdef{test-rely}
with the fact that $\Variable{rx}$ and $\Variable{ry}$ are local variables and hence unmodified by interference,
\begin{align}
  r \defs \Rel{\Deref{f} \subseteq \Deref{f}' \land roots(\Deref{f}') \subseteq roots(\Deref{f}) \land \Deref{rx}' = \Deref{rx} \land \Deref{ry}' = \Deref{ry}}. \labeldef{test-rely-local}
\end{align}
Taking $\Ex{f_1}$ to be the value of $\Deref{f}$ before guard evaluation, first we show,
\begin{align}
  \TripleV{\Set{f_1 \subseteq \Deref{f} \land roots(\Deref{f}) \subseteq roots(f_1)}}{\rely{r} \together \Expr{\Deref{(\Array{f}{\Deref{rx}})}}{k}}{\Set{\exists f_2 \spot \Array{\Ex{f_2}}{\Deref{rx}} = k \land f_1 \subseteq f_2 \subseteq \Deref{f} \land roots(\Deref{f}) \subseteq roots(f_2) \subseteq roots(f_1)}} \labelprop{eval-f-rx}
\end{align}
by \reflem{post-array-invar-index-deref} with 
\begin{align}
  P\,\Ex{k} \defs \Set{\exists f_2 \spot \Array{\Ex{f_2}}{\Deref{rx}} = k \land f_1 \subseteq f_2 \subseteq \Deref{f} \land roots(\Deref{f}) \subseteq roots(f_2) \subseteq roots(f_1)}
\end{align}
because $\Set{f_1 \subseteq \Deref{f} \land roots(\Deref{f}) \subseteq roots(f_1) \land \Deref{(\Array{f}{\Deref{rx}})} = k} \subseteq P\,\Ex{k}$ 
taking $f_2$ as $\Deref{f}$ for the existential
and $P\,\Ex{k}$ is stable under $r$ \refdef{test-rely-local}.

For the binary expression $\Deref{rx} \neq \Deref{(\Array{f}{\Deref{rx}})}$, the value of $\Deref{rx}$ is invariant under $r$ 
and hence by \reflem{post-expr-invariant}
\begin{align}
  \Triple{p}{\Expr{\Deref{rx}}{k_1}}{p \inter \Set{\Deref{rx} = k_1}} \labelprop{deref-rx}
\end{align}
Applying \reflem{post-binary} with binary operator $\neq$ using \refprop{deref-rx} and \refprop{eval-f-rx} we establish the following.
\begin{align}
  \EstablishExprV{\Set{f_1 \subseteq \Deref{f} \land roots(\Deref{f}) \subseteq roots(f_1)}}{r}{\Deref{rx} \neq \Deref{(\Array{f}{\Deref{rx}})}}{\True}{\Set{\exists f_2 \spot \Deref{rx} \neq \Array{\Ex{f_2}}{\Deref{rx}} \land f_1 \subseteq f_2 \subseteq \Deref{f} \land roots(\Deref{f}) \subseteq roots(f_2) \subseteq roots(f_1)}} \labelprop{rx-neq-f-rx}
\end{align}
For the case where the guard evaluates to $\Ex{\True}$, (i.e.\ taking $\Ex{k}$ to be $\Ex{\True}$)
the postcondition becomes,
\begin{align}&
  ~\Ex{\exists f_2 \spot (\Deref{rx} \neq \Array{\Ex{f_2}}{\Deref{rx}} \land f_1 \subseteq f_2 \subseteq \Deref{f} \land roots(\Deref{f}) \subseteq roots(f_2) \subseteq roots(f_1))}
 \Implies
  ~\Ex{\exists \Ex{f_2} \spot (\Deref{rx} \not\in roots(f_2) \land roots(\Deref{f}) \subseteq roots(f_2))}
 \Implies
  ~\Ex{\Deref{rx} \not\in roots(\Deref{f})}.
\end{align}
Similar conditions hold for the expression $\Deref{ry} \neq \Deref{(\Array{f}{\Deref{ry}})}$,
and hence one can conclude both,
\begin{align*}
  \EstablishExpr{\Set{f_1 \subseteq \Deref{f} \land roots(\Deref{f}) \subseteq roots(f_1)}}{r}{\Deref{rx} \neq \Deref{(\Array{f}{\Deref{rx}})}}{\True}{\Set{\Deref{rx} \not\in roots(\Deref{f})}} \\
  \EstablishExpr{\Set{f_1 \subseteq \Deref{f} \land roots(\Deref{f}) \subseteq roots(f_1)}}{r}{\Deref{ry} \neq \Deref{(\Array{f}{\Deref{ry}})}}{\True}{\Set{\Deref{ry} \not\in roots(\Deref{f})}} 
\end{align*}
and then using \reflem{post-disjoin} one can compose these results to give,
\begin{align*}
  \EstablishExprV{\Set{f_1 \subseteq \Deref{f} \land roots(\Deref{f}) \subseteq roots(f_1)}}{r}{\Deref{rx} \neq \Deref{(\Array{f}{\Deref{rx}})} \lor \Deref{ry} \neq \Deref{(\Array{f}{\Deref{ry}})}}{\True}{\Set{\Deref{rx} \not\in roots(\Deref{f}) \lor \Deref{ry} \not\in roots(\Deref{f})}} 
\end{align*}
and hence $\Set{\Deref{rx} \not\in roots(\Deref{f}) \lor \Deref{ry} \not\in roots(\Deref{f})}$ becomes the precondition for the body of the loop.
This case is straightforward because the boolean expression evaluating to $\Ex{\True}$ is stable under the rely $r$ \refdef{test-rely-local}.

The case where the guard evaluates to $\Ex{\False}$ is more subtle
because neither $\Set{\Deref{rx} \in roots(\Deref{f})}$ nor $\Set{\Deref{ry} \in roots(\Deref{f})}$ is stable under the rely \refdef{test-rely}.
For $\Ex{k}$ as $\Ex{\False}$ in \refprop{rx-neq-f-rx},
the postcondition becomes, 
\begin{align}&
  ~\Ex{\exists f_2 \spot (\Deref{rx} = \Array{\Ex{f_2}}{\Deref{rx}} \land f_1 \subseteq f_2 \subseteq \Deref{f} \land roots(\Deref{f}) \subseteq roots(f_2) \subseteq roots(f_1))}
 \Implies
  ~\Ex{\exists f_2 \spot (\Deref{rx} \in roots(f_2) \land roots(f_2) \subseteq roots(f_1))}
 \Implies
  ~\Ex{\Deref{rx} \in roots(f_1)}.
\end{align}
Similar conditions hold for $\Ex{ry}$ and hence one can deduce both,
\begin{align*}
  \EstablishExpr{\Set{f_1 \subseteq \Deref{f} \land roots(\Deref{f}) \subseteq roots(f_1)}}{r}{\Deref{rx} = \Deref{(\Array{f}{\Deref{rx}})}}{\True}{\Set{f_1 \subseteq \Deref{f} \land \Deref{rx} \in roots(f_1)}} \\
  \EstablishExpr{\Set{f_1 \subseteq \Deref{f} \land roots(\Deref{f}) \subseteq roots(f_1)}}{r}{\Deref{ry} = \Deref{(\Array{f}{\Deref{ry}})}}{\True}{\Set{f_1 \subseteq \Deref{f} \land \Deref{ry} \in roots(f_1)}}
\end{align*}
and then using \reflem{post-conjoin} one can compose these results to show the boolean expression.
\begin{align*}
  \EstablishExpr{\Set{f_1 \subseteq \Deref{f} \land roots(\Deref{f}) \subseteq roots(f_1)}}{r}{\Deref{rx} = \Deref{(\Array{f}{\Deref{rx}})} \land \Deref{ry} = \Deref{(\Array{f}{\Deref{ry}})}}{\True}{p_f} ,
\end{align*}
where 
\begin{align}
  p_f \defs \Set{f_1 \subseteq \Deref{f} \land \Deref{rx} \in roots(f_1) \land \Deref{ry} \in roots(f_1)},  \labeldef{pf}
\end{align}
or equivalently using \refprop{expr-not} and de Morgan's law \refprop{de-morgan1}, 
\begin{align*}
  \EstablishExpr{\Set{f_1 \subseteq \Deref{f} \land roots(\Deref{f}) \subseteq roots(f_1)}}{r}{\Deref{rx} \neq \Deref{(\Array{f}{\Deref{rx}})} \lor \Deref{ry} \neq \Deref{(\Array{f}{\Deref{ry}})}}{\False}{p_f},
\end{align*}
and hence $p_f$ becomes the condition on exit from the loop.
For any elements $v$ and $u$, 
\begin{align}
  v \in roots\,f \land u \in roots\,f \implies ((v \equiv_f u) \iff (v = u)) \labelprop{equiv-for-roots}
\end{align}
and hence for the forest $\Ex{f_1}$ we have the property $\Ex{(\Deref{x} \equiv_{f_1} \Deref{y}) \equiv (\Deref{rx} = \Deref{ry})}$ as follows. 
\begin{align*}&
  \Ex{\Deref{x} \equiv_{f_1} \Deref{y}} 
 \Equiv*[by transitivity as from the invariant \refprop{test-inv} $\Deref{x} \equiv_{f_1} \Deref{rx}$ and $\Deref{y} \equiv_{f_1} \Deref{ry}$]
  \Ex{\Deref{rx} \equiv_{f_1} \Deref{ry}}
 \Equiv*[by \refprop{equiv-for-roots} as $\Deref{rx}$ and $\Deref{ry}$ are both roots of $\Ex{f_1}$]
  \Ex{\Deref{rx} = \Deref{ry}} 
\end{align*}
If we let $\Ex{f}$ stand for the forest on entry to $test$,
for the the first half of the postcondition \refprop{test-post} of $test$ 
using the above property and $\Ex{\Deref{f} \subseteq f_1}$ (because the forest only grows \refdef{f-grows}), we have,
\begin{align*}
  \Ex{(\Deref{x} \equiv_{\Deref{f}} \Deref{y})} \implies & 
  ~\Ex{(\Deref{x} \equiv_{f_1} \Deref{y})} \\
  = &~\Ex{(\Deref{rx} = \Deref{ry})}
\end{align*}
For the second half of the postcondition of $test$ \refprop{test-post}, 
if on termination of the loop $\Deref{rx} = \Deref{ry}$ and $\Deref{f}'$ is the final value of the forest,
then from the invariant \refprop{test-inv} we have $\Deref{x} \equiv_{\Deref{f}'} \Deref{rx}$ and $\Deref{y} \equiv_{\Deref{f}'} \Deref{ry}$
and hence by transitivity of the equivalence relation, $\Deref{x} \equiv_{\Deref{f}'} \Deref{y}$, that is,
\begin{align*}
  \Ex{(\Deref{rx} = \Deref{ry}) \implies (\Deref{x} \equiv_{\Deref{f}'} \Deref{y})}.
\end{align*}
Combining the above two sets of reasoning we get
\begin{align*}
  \Ex{((\Deref{x} \equiv_{\Deref{f}} \Deref{y}) \implies (\Deref{rx} = \Deref{ry})) \land ((\Deref{rx} = \Deref{ry}) \implies (\Deref{x} \equiv_{\Deref{f}'} \Deref{y}))}
\end{align*}
which after the assignment $\Assignment{t}{(\Deref{rx} = \Deref{ry})}$ 
gives the postcondition of $test$ \refprop{test-post}.

For those familiar with the Fischer-Galler forest-based approach,
Jayanti and Tarjan \cite{JayantiTarjan21} have examined variations of 
concurrent algorithms for handling equivalence relations represented as forests
and their computational complexity.
They consider path compression techniques used to improve the performance of the approach.
The path compression technique that reduces the path from an element to its root by replacing a 
reference to its parent by a reference to its grandparent while traversing a path to the root,
maintains the rely conditions required for $test$.

\section{Conclusion}\labelsect{conclusion}

In shared-memory concurrent programs, expression evaluation is not free from interference by parallel threads; thus, following the evaluation of an expression $\Ex{e}$ to a value $\Ex{k}$, one cannot necessarily assume that the expression $\Ex{e}$ equals $\Ex{k}$ in the state following the expression evaluation
nor indeed that $\Ex{e}$ evaluated to $\Ex{k}$ in any single state during its evaluation
(i.e.\ that the evaluation was atomic).

Modelling expression evaluation and assignment commands as atomic,
an approach taken in many examples of earlier work
\cite{Jones81d,Jones83a,Jones83b,Stolen90,stolen1991method,XuRoeverHe97,Dingel02,PrensaNieto03,Wickerson10-TR,DBLP:conf/esop/WickersonDP10,SchellhornTEPR14,Sanan21} 
simplifies writing a semantics of the programming language,
but is unrealistic for arbitrary expressions and assignments. 
Justifications for these atomicity assumptions often appeal to syntactic constraints 
that expressions and assignments contain at most one critical reference.
The validity of these syntactic constraints is not, 
however, justifiable with respect to the restricted programming language semantics used in those approaches
and thus breaks the formal chain of reasoning. 
Furthermore,
copying test expressions from conditionals or while loops as pre conditions for the sub-commands is invalid in the presence of interference 
even if there is only one occurrence of a shared variable in the expression.
Rely-guarantee rules resolve these issues but most previous soundness proofs have been based on 
operational semantic descriptions.
Coleman and Jones~\cite{CoJo07} provide a fine-grained operational semantics that is similar to the semantic approach used here, for example expression evaluation and assignment commands are not assumed to be atomic, but the inference rules they develop for commands containing expression evaluation are more restrictive;
they can be seen as special cases of our more general laws.

Using an algebraic presentation of the rely-guarantee theory,
we have presented formal proofs of refinement laws
with respect to a programming language with minimal atomicity assumptions (just atomic access to individual l-values).
Moreover our laws are semantic rather than syntactic leading to further generalisation.
We have presented compositional laws for reasoning about expression evaluation 
that do not require syntactic single critical reference constraints;
our laws have been used to justify proof obligations for refinement laws for 
conditionals, assignment commands and while-loops.

The genesis of the approach presented in this paper was 
the realisation that boolean expressions formed from conjunctions and disjunctions
could be treated compositionally as in \reflem{post-conjoin} and \reflem{post-disjoin}.
From there it was a matter of applying our program algebra to devise the more general laws for binary and unary expressions,
from which the laws for conjunction and disjunction are now derived.
Our rules avoid the necessity to rewrite the program so that 
all expressions and assignments have at most one critical reference
(the informal syntactic assumption used in most  approaches to shared-variable concurrency).
Our rely-guarantee theory has been formalised within the Isabelle/HOL theorem prover \cite{IsabelleHOL}
and proofs of all the above theorems have been performed within our theory.

By using a fine-grained semantics for expressions (instead of assuming they are atomic), 
we have been able to derive a compositional set of rules for reasoning about arbitrary side-effect-free expressions
in the context of interference bounded by a rely condition.
Overall this makes reasoning about concurrent programs simpler and 
avoids introducing unnecessary additional local variables
and, significantly in the context of trusted systems,
maintains the formal chain of reasoning.
Our theory depends crucially on using a rely condition to bound interference 
and the concept of stability from standard rely guarantee theory.

\begin{acks}
Our joint work is supported by the 
\grantsponsor{ARC}{Australian Research Council}{https://www.arc.gov.au/grants/discovery-program}
under their Discovery Program Grant 
No.~\grantnum{ARC}{DP190102142}.
Hayes and Meinicke are also supported by funding from the 
\grantsponsor{DSTG}{Department of Defence, administered through the Advanced Strategic Capabilities Accelerator}{TODO}
grant \grantnum{DSTG}{Verifying Concurrent Data Structures for Trustworthy Systems}.
Jones is also supported by the
\grantsponsor{Leverhulme}{Leverhulme Trust}{TODO}
No.~\grantnum{Leverhulme}{RPG-2019-020}
and would like to add thanks to the stimulus of the  {\em  Big Specification} programme at the {\em Isaac Newton Institute}.
\end{acks}

\bibliographystyle{ACM-Reference-Format}
\citestyle{acmnumeric}
\bibliography{ms}

\end{document}